\documentclass[12pt]{cernrep}
\usepackage{graphicx,wrapfig,epsfig,epsf}
\usepackage{amsmath,amssymb,enumerate}
\usepackage{hhline,multirow}

\newcommand{\sNN}{s_{\mbox{\tiny NN}}}
\newcommand{\yNN}{\Delta y}
\newcommand{\EN}{E_{\mbox{\tiny N}}}

\def\gsim{~\raise0.3ex\hbox{$>$\kern-0.75em\raise-1.1ex\hbox{$\sim$}}~}
\def\lsim{~\raise0.3ex\hbox{$<$\kern-0.75em\raise-1.1ex\hbox{$\sim$}}~}

\newcommand{\be}{\begin{eqnarray}}
\newcommand{\ee}{\end{eqnarray}}

\def\Z0{$\mathrm{Z^0}$}

\def\pp{p+p\,}
\def\pA{p+A\,}
\def\Ap{A+p\,}
\def\AA{A+A\,}
\def\PbPb{Pb+Pb\,}
\def\pPb{p+Pb\,}
\def\dPb{d+Pb\,}
\def\dAu{d+Au\,}
\def\dA{d+A\,}
\def\AuAu{Au+Au\,}

\begin{document}

\title{Proton-Nucleus Collisions at the LHC: \\ Scientific Opportunities and Requirements}

\def\compostella{1}
\def\annecy{2}
\def\nikhef{3}
\def\jussieux{4}
\def\fermilabth{5}
\def\cernacc{6}
\def\neviscol{7}
\def\cernep{8}
\def\bcn{9}
\def\cea{10}
\def\jlab{11}
\def\basel{12}
\def\lbnl{13}
\def\louvain{14}
\def\bnl{15}
\def\kfki{16}
\def\psu{17}
\def\davis{18}
\def\muns{19}
\def\cernth{20}
\def\mit{21}
\def\stpeter{22}

\author{
{\bf Editor:} C.~A.~Salgado$^{\compostella}$\\
{\bf Authors:} 
J. Alvarez-Mu\~niz$^{\compostella}$,
F.~Arleo$^{\annecy}$,
N.~Armesto$^{\compostella}$,
M.~ Botje$^{\nikhef}$,
M.~Cacciari$^{\jussieux}$,
J.~Campbell$^{\fermilabth}$,
C.~Carli$^{\cernacc}$,
B.~Cole$^{\neviscol}$,
D.~D'Enterria$^{\cernep,\bcn}$, 
F.~Gelis$^{\cea}$,
V.~Guzey$^{\jlab}$, 
K.~Hencken$^{\basel}\footnote{~~Current address:
ABB Switzerland Ltd., Corporate Research, Baden-D\"attwil, Switzerland}$,
P.~Jacobs$^{\lbnl}$,
J.~M.~Jowett$^{\cernacc}$, 
S.~R.~Klein$^{\lbnl}$,
F.~Maltoni$^{\louvain}$,
A.~Morsch$^{\cernep}$, 
K.~Piotrzkowski$^{\louvain}$,
J.~W.~Qiu$^{\bnl}$,
T.~Satogata$^{\bnl}$,
F.~Sikler$^{\kfki}$, 
M.~Strikman$^{\psu}$,
H.~Takai$^{\bnl}$, 
R.~Vogt$^{\lbnl,\davis}$,
J.~P.~Wessels$^{\cernep,\muns}$,
S.~N.~White$^{\bnl}$, 
U.~A.~Wiedemann$^{\cernth}$,
B.~Wyslouch$^{\mit}$\footnote{~~On leave of absence, Massachusetts Institute of Technology, Cambridge, MA 02139, USA},
M.~Zhalov$^{\stpeter}$
}

\institute{
$^{~\compostella}$
Departamento de F\'{\i}sica de Part\'{\i}culas and IGFAE, U. Santiago de Compostela, Galicia, Spain\\
$^{~\annecy}$
LAPTH, Universit\`e de Savoie et CNRS, Annecy-le-Vieux Cedex, France \\
$^{~\nikhef}$
NIKHEF, Amsterdam, The Netherlands\\
$^{~\jussieux}$
LPTHE,  Universit\'e Pierre et Marie Curi\'e (Paris 6), France\\
$^{~\fermilabth}$
Theoretical Physics Department, Fermilab, Batavia, IL, USA\\
$^{~\cernacc}$
Beams Department, CERN, Geneva, Switzerland \\
$^{~\neviscol}$
Nevis Laboratories, Columbia University, New York, NY, USA\\
$^{~\cernep}$
Physics Department, 
Experimental Division, CERN, Geneva, Switzerland\\
$^{~\bcn}$
ICREA, ICC-UB, Univ. de Barcelona, 08028 Barcelona, Catalonia\\
$^{~\cea}$
IPTh, CEA/DSM/Saclay, 91191, Gif-sur-Yvette Cedex, France\\
$^{~\jlab}$ 
Jefferson Lab, Newport News, VA, USA\\
$^{~\basel}$
Institut f\"{u}r Physik, Universit\"{a}t  Basel,  Switzerland \\
$^{~\lbnl}$
Nuclear Science Division, Lawrence Berkeley National Laboratory, Berkeley, CA, USA\\
$^{~\louvain}$
Universit\'{e} Catholique de Louvain, Louvain-la-Neuve, Belgium \\
$^{~\bnl}$
Physics Department, Brookhaven National Laboratory, Upton, NY 11973, USA\\
$^{~\kfki}$
KFKI Research Institute for Particle and Nuclear Physics, Budapest, Hungary\\
$^{~\psu}$
Department of Physics, Pennsylvania State University, USA\\
$^{~\davis}$
Physics Department, University of California at Davis, Davis, CA, USA\\
$^{~\muns}$
Institut  f\"{u}r  Kernphysik,  Universit\"at Muenster, D-48149 Muenster, Germany\\
$^{~\cernth}$
Physics Department, Theory Division, CERN, Geneva, Switzerland\\
$^{~\mit}$
LLR Ecole Polytechnique, 91128 Palaisseau Cedex, France\\
$^{~\stpeter}$
St. Petersburg Nuclear Physics Institute, Gatchina, Russia\\
}

\maketitle 

\begin{flushright}
\vspace{-19cm}
CERN-PH-TH/2011-119\\
LHC-Project-Report-1181\\
\vspace{18cm}
\end{flushright} 

\begin{abstract}
Proton-nucleus (p+A) collisions have long been recognized as a crucial
component of the physics programme with nuclear beams at high
energies, in particular for their reference role to interpret and
understand nucleus-nucleus data as well as for their potential to
elucidate the partonic structure of matter at low parton fractional
momenta (small-$x$). Here, we summarize the main motivations that make
a proton-nucleus run a decisive ingredient for a successful heavy-ion
programme at the Large Hadron Collider (LHC) and we present unique
scientific opportunities arising from these collisions.  We also
review the status of ongoing discussions about operation plans for
the \pA mode at the LHC.
\end{abstract}

\newpage
\tableofcontents


\newpage
\section{EXECUTIVE SUMMARY}
\label{sec:summary}
Heavy-ion physics is an integral part of the baseline experimental
programme of the CERN Large Hadron Collider (LHC). After normal
operations have been established, the LHC is running for about 8
months per year with proton beams and for one month per year with
nuclear beams. Three of the four experiments (ALICE, ATLAS and CMS)
participate in the LHC nuclear beam programme\footnote{Although LHCb has not so far
considered running with nuclear beams, there is in principle no
technical difficulty preventing this experiment from doing so. Their
excellent detection capabilities at forward rapidities would be very
useful for  proton-nucleus measurements.}. So far, only collisions
with Pb nuclei are firmly scheduled, while operational plans for
proton-nucleus (\pA) collisions are still preliminary. However, the LHC
is a versatile hadron collider that allows, in principle, the
collision of asymmetric (A+B) nuclear beams. All three LHC experiments
have included \pA collisions in their physics performance studies and
have discussed their importance.

The proton-nucleus programme serves a dual purpose. It provides, on
the one hand, baseline measurements for the nucleus-nucleus
program. Experience from previous heavy ion programs (CERN SPS, RHIC)
shows that a \pA baseline is essential for the interpretation of some of
the main discoveries (e.g. $J/\psi$-suppression, jet quenching,
...). This document identifies an analogous need for \pA collisions at
the LHC. A \pA programme also offers unique
possibilities for specific investigations in various domains of
Quantum Chromodynamics (QCD).

This document presents an updated description of: i) the
accelerator issues for collision of asymmetric systems at the LHC; ii)
the uncertainties in nuclear parton distribution functions and
benchmark cross section for hard processes; iii) the new opportunities
made available to study parton saturation, ultra-peripheral collisions and
measurements, which are of interest to astrophysics; and iv) the experimental issues
related to the special conditions of the \pA run.

The main conclusions of this document are the following:

\begin{itemize}

\item Preliminary considerations indicate the feasibility to run the LHC
in \pA mode without major modifications of the machine. We consider
here a canonical situation in which the energy of the \pPb run
($\sqrt{s}=8.8$ TeV) corresponds to the charge-over-mass ratio scaling
with respect to the proton top LHC energy. The estimated luminosity is
$L=10^{29}$cm$^{-2}$s$^{-1}$ for \pPb collisions. Asymmetric
collisions imply also rapidity shifts with respect to the \AA and \pp
systems. This effect can be reduced by colliding
deuterons with nuclei instead of protons. The realisation of the d+A
mode, however, needs significant hardware modifications of the
injector chain.

\item The knowledge of nuclear parton distributions is deficient for the
kinematics accessible at the LHC. This has negative consequences for
the interpretation of the \AA data and a \pA run is indispensable
for benchmarking: For most of the duration of the LHC
programme, a \pA run will be the only experimental possibility to reduce
systematic uncertainties arising from yet unmeasured parton
distributions. In this document we study possible constraints
from different processes. Assuming a running time of $10^6\, s$ the
corresponding integrated luminosity of 0.1 pb$^{-1}$ will 
make the measurements considered in this document feasible. A significantly
smaller integrated luminosity will compromise the performance for
several observables. On the contrary, an increase in luminosity
by a factor of 10 will be beneficial for observables with the smallest cross
sections, especially those involving high-$p_T$ photons and heavy
bosons.

\item \pA collisions at the LHC offer unique possibilities for the study of
small-$x$ physics with nuclear targets, extending the kinematically
accessible regime by several orders of magnitude in $x$ and $Q^2$ from those
presently available. This provides an excellent opportunity for
the study of the saturation of partonic densities. Other
physics opportunities include ultra-peripheral (electromagnetic)
collisions and measurements of cross sections of interest for cosmic
ray physics.

\item The LHC experiments have proven capabilities to exploit the physics
opportunities of a \pA run at the quoted luminosity. For some
observables, running \pPb at the same nucleon-nucleon centre-of-mass
energy as \PbPb would have the
advantage of reducing systematic uncertainties for benchmark
perturbative processes. Further reduction of the uncertainties, for
the case of rapidity asymmetric detectors, necessitates operation of
both \pA and \Ap modes. For nuclear PDF and small-$x$ studies, the
largest possible energy is preferred. This document provides important
arguments for the timely scheduling of \pA runs.

\end{itemize}

 \clearpage

\newpage
\section{INTRODUCTION}
\label{pA_SEC:intro}

With the LHC, high-energy nuclear collisions have reached the TeV scale for
the first time. This has opened up a discovery regime, which is
currently being vigorously
explored exploiting the first long run with \PbPb collisions at $\sqrt{s_{\rm
NN}}$= 2.76 TeV. The corresponding first results on multiplicities,
elliptic flow, jet quenching and other observables have been
published~\cite{Aamodt:2010pb,Aamodt:2010pa,Aamodt:2010jd,Aad:2010bu,Chatrchyan:2011sx}. The
top LHC energy of $\sqrt{s_{\rm NN}}$=5.5 TeV exceeds that of the
Relativistic Heavy Ion Collider (RHIC) by almost a factor of 30.  The
increase in centre-of-mass energy will be even larger for \pPb collisions
($\sqrt{s_{\rm NN}}$=8.8~TeV).  This large jump in energy translates
into a kinematical reach in
Bjorken-$x$ and virtuality $Q^2$ that is several orders of magnitude beyond
 that achieved in all other previous experiments with nuclear
collisions.

 Based on the current understanding of nucleus-nucleus collisions,
there are in particular two classes of questions where the extended
kinematic reach of the LHC is expected to give access to qualitatively
new phenomena.  i) Hard probes are produced at unprecedented rate
at the LHC. The established strong sensitivity of these hard probes
provides a promising
and diverse method for a detailed characterization of the properties
of the produced dense QCD matter; ii) Much smaller momentum fractions
$x$ become relevant for particle production. On the one hand, the
large parton densities at small-$x$ are expected to make the system
initially produced in the collision denser, hotter and thus longer
lived. As a result, a characterization of parton distributions at
small-$x$ is important for the understanding of the initial conditions from
which dense QCD matter emerges --- as also demonstrated by the first particle 
multiplicity data \cite{Aamodt:2010pb} showing striking scaling features with 
respect to the smaller energy (RHIC) data.
On the other hand, the dense initial
partonic system is of interest in its own, since one expects to access
with increasing $\sqrt{s}$ a novel high density regime of QCD in which
parton distributions are saturated up to perturbatively large
virtualities.  As discussed in this document, measurements of proton-nucleus
collisions are crucial for exploiting the opportunities of these two
classes of measurements at the LHC.

Normalization runs with proton-nucleus (p+A) collisions have long been
recognized as a crucial component of the LHC heavy-ion program. More
generally, this is because the characterization of signatures of the
QCD matter created in heavy ion collisions relies on benchmark
processes with elementary collision partners, in which final-state
medium effects such as collective phenomena are largely absent.  To some
extent, such benchmarking can be done with data from proton-proton
collisions which characterize production processes in the
absence of both initial and final state medium effects. However, to
disentangle the initial state effects from those final state effects
which characterize the properties of the produced dense QCD matter,
proton-nucleus collisions are crucial. The absence of strong final
state medium effects in proton-nucleus collisions provides a unique
opportunity for characterizing the nuclear dependence of parton
distribution functions at a hadron collider. This benchmarking of the
initial conditions is of particular importance for heavy-ion
collisions at the LHC, since e$+$A DIS data will not be available
any time soon for a large range
of the kinematically relevant small-$x$ values. Beyond benchmarking,
proton-nucleus collisions are also expected to provide access to qualitatively
new features of the small-$x$ structure of matter.

Since the last extensive discussions of proton-nucleus collisions at
the
LHC \cite{Accardi:2003be,Accardi:2003gp,Bedjidian:2003gd,Arleo:2003gn},
significant experimental and theoretical developments have occurred in
this area. In particular, the technical challenge of running a hadron
collider in an asymmetric mode has been mastered at RHIC, where
deuterium-gold measurements have provided the decisive benchmark
experiments for discoveries in the corresponding nucleus-nucleus
program.  The extrapolation of these results allow us now to substantiate
statements about the required luminosity and experimental coverage
for a successful proton-nucleus normalization run at the LHC. Further,
recent theoretical developments in studying the energy evolution of
Quantum Chromodynamics at high parton densities and small-$x$ have
added significantly to the physics case for a proton-nucleus run at
the LHC. In fact, a growing physics community has started
to work towards a dedicated electron-nucleus collider for elucidating
the small-$x$ structure of matter~\cite{LHeC,EIC}. The question arises to
what extent proton-nucleus runs at the LHC can help to prepare and
complement such a future large scale project.

The purpose of this document is to provide a brief update of our current
understanding of how the LHC could operate in \pA mode, how a
proton-nucleus run could be exploited to  optimally support a
successful heavy-ion programme at the LHC, and which additional
scientific opportunities arise from proton-nucleus collisions at the
LHC.


\section{THE LHC AS A PROTON-NUCLEUS COLLIDER}
\label{sec:LHCmachine}

Although proton-nucleus collisions (\pA) at the LHC had been discussed
in the physics community for some years, they were not formally
included in the initial ``baseline'' LHC machine
design~\cite{ref:LHCDesignReport}, which included only symmetric \pp
and \PbPb collisions. The \pA mode of operations (as well as
collisions between lighter ions) was considered as an option,
to be studied and implemented later.  The rationale for this at the
time was twofold:
\begin{itemize}

\item The need to focus available resources on what was needed for the
startup of the LHC.

\item The baseline modes of operation are the most difficult in many respects.

\end{itemize}
 
With the LHC starting to explore
 the \pp and \PbPb physics programmes, a need to investigate
the feasibility of the \pA mode of operation of the LHC appears.
Meanwhile some concerns  have emerged from the
experience gained at RHIC. In this document we have tried to respond to these
concerns with a preliminary study of the feasibility and potential
performance.

\subsection{RHIC Experience   }\label{sec:RHICexperience}

 RHIC operated with deuteron-gold collisions (\dAu) in 2003 and 2008 \cite{RHICdAu1,
 RHICdAu2}. The
 2003 run was the first for an asymmetric hadron collider, and faced
 several challenges. Some relevant to the LHC proton-ion programme include
 multi-species injector performance, setup time constraints, injection with
 equal rigidity vs. equal revolution frequency between the two beams, and
 collision geometries of dissimilar species. At RHIC a proton-gold run was considered but
required either large deviations in the arcs or
 movement of the common DX magnets. So, the option of a
 more symmetric charge-over-mass ratio as in a \dAu colliding system was adopted.

 Careful attention should be given to dual-species injector performance and
 reliability, particularly since any bottleneck will significantly impact
 the performance of short LHC \pPb runs. Injector emittance and
 intensity development ultimately limited RHIC performance during the 2003
 \dAu run. Later improvements in injector Au performance for dedicated \AuAu
 operations provided the basis for the 2008 \dAu run, which delivered six
 times the integrated luminosity of that in 2003.

RHIC \dAu took 18~days of development to first collisions, and an
additional 20~days
to the start of physics development.  This time included the
development of new acceleration/\(\beta\)-squeeze ramps (see below) and
about 6~days of detector setup operations. The
LHC \pPb setup will likely be
shorter since LHC \pp and \PbPb will precede \pPb operations.

Early in the RHIC 2003
\dAu run, injection and ramping setup were changed from the same
magnetic rigidity in both rings to the same RF frequency in both
rings---this was necessary to avoid beam-beam modulation and serious beam
losses (up to 50~\% of stored beam) during injection and ramping, even
with interaction region beam separations of 10 sigma.  For the LHC,
beam-beam compensation  and transverse dampers should be investigated,
perhaps with modifications, to control emittance growth and beam loss
from this mechanism.

The collision geometry of the LHC for \pPb should be studied for
asymmetries, in conjunction with the LHC experiments.
The RHIC \dAu run
required swaps of power supply shunts on DX dual-ring magnets to satisfy
the orbit geometry requirements of different
\(Z/A\)
species in each ring.
This created a \(\QTY{1}{\mu rad}\) collision angle within the
experiments, marginally affecting detector acceptances.

\subsection{Injector chain for proton-ion or deuteron ion operation of the LHC}\label{sec:LHCInjectors}
 
For the reasons given above, only
preliminary considerations on LHC filling for \pA or \dA
ope\-ra\-tion have been made.
The injector  chains for protons and ions are distinct at the low
energy end, i.e., the initial Linac and the first synchrotron.
Clearly, as much as possible of
the existing LHC injector chains  must be used to
allow  proton (deuteron)-ion operation of the LHC at a reasonable
cost.

We assume that one of the two LHC rings will be filled with ions
featuring the nominal bunch pattern for ion-ion operation.  Furthermore,
we assume that the second LHC ring must be filled with protons or
deuterons in
the \emph{same nominal ion} bunch pattern.  The ``lazy'' solution, to
fill the other ring with the usual \pp bunch pattern and to
have at least some encounters at the interaction points (but some ion
bunches may never collide with a proton bunch), would have many
disadvantages and is not envisaged.

\subsubsection{Injector chain for proton-ion operation}\label{sec:injp}

Assuming that one LHC ring is filled with a nominal Pb-ion beam
(bunch
population, emittances, filling pattern)   and that
the proton bunches have the same geometric beam sizes and bunch pattern,
an
intensity of the order of \(10^{10}\)~protons per bunch is needed to reach the
required luminosity of the order of \(10^{29}\rm cm^{-2}s^{-1}\).

The LHC ion ring will be filled using the standard ion injector chain in
the standard way (yielding the nominal LHC ion bunch pattern).  The
problem consists in finding a scheme that allows  the LHC proton
ring to be filled with protons in the same bunch pattern as the ions.
Two distinct schemes have
been identified (although many more probably exist).  Each of them uses the LHC
proton injector chain but applies longitudinal
gymnastics that are very different from nominal proton filling.

\begin{itemize}

\item The first scheme is based on the experience with generation of
LHC (proton) pilot bunches in the PS Booster.  Every Booster ring
provides one bunch per cycle and each of these bunches corresponds to
one LHC bunch (no bunch splitting along the rest of the chain).  These
bunches are injected into adjacent PS buckets.  The harmonic number
of the PS (\(h=16\) may be a good choice)
is chosen so that the bunch spacing
is sufficient for the PSB recombination line kicker.  The PS harmonic
number must be increased gradually to \(h = 21\) to obtain the right
100~ns bunch spacing and the 40~MHz and 80~MHz RF systems must be used
to
shorten the bunches before ejection towards the SPS.  The four bunches
provided per PS proton cycle correspond to the four ion bunches  per
PS ion cycle.  The rest of the LHC injector chain is very similar
(accumulation of up to 13~injections on an SPS low energy plateau,
acceleration and transfer to the LHC) for proton and ion filling and
even slightly simpler (no stripping in the transfer line between PS and
SPS and, thus, a higher magnetic field in the SPS, no ``fixed''
frequency non-integer harmonic acceleration needed for protons).

\item The second scheme aims at faster filling of the LHC proton ring.
Every bunch provided by the PS Booster corresponds to one LEIR/PS ion
cycle and, thus, has to provide four LHC proton bunches.  This scheme
requires more elaborate longitudinal gymnastics than the
first one  (but they are still simpler than the
gymnastics applied routinely in operations to provide, e.g., the beam
for LHC \pp operation).

\end{itemize}

At a first glance, the proton beam needed for ion-proton operation of
the LHC can be provided by the injector chain at a reasonable cost and
without major hardware upgrades.

\subsubsection{Injector Chain for Deuteron-Ion Operation of the LHC  }\label{sec:injd}

Schemes for filling one LHC ring with deuterons have been given only
very preliminary consideration.  In principle, one could imagine one of
the following two scenarios:

\paragraph{ Deuterons via the ion injector chain (Linac 3 and LEIR):}
In
order to provide both ions and deuterons via the ion injector chain,
sufficiently fast switching between these two species at the low energy
part of Linac3 would be required.  Furthermore the radio-frequency
quadrupole (RFQ) used for ions would not perform well for deuterons.
Thus, a dedicated deuteron RFQ, adjusted to higher input particle
velocity and voltage for extraction from the source, would be necessary.
In summary it is clear that a dedicated deuteron source and RFQ and a
switchyard allowing switching between ions and deuterons is the minimum
requirement.

\paragraph{ Deuterons via the proton injector chain (Linac~2 PSB):}
Linac4 will replace the present Linac2 as PS Booster injector from 2015 on, so this
option will not be available after that time.
The  PS Booster will then be converted for $\mathrm{H}^-$ charge exchange injection
so deuterons for the LHC would have to be produced from $\mathrm{D}^-$ injection.
 Whereas the simple   drift tube structure of Linac2 could,
in principle, accelerate $\mathrm{D}^-$ (with a
velocity half that of protons), this is not the case
for Linac4 which consists of three different accelerating
structures.  A dedicated $\mathrm{D}^-$ source and RFQ would be
needed in any case and represents a significant  investment.

Using Linac2 for $\mathrm{D}^-$ acceleration would have limited impact on
proton beams for other facilities.

These very preliminary investigations lead to the conclusion that the
injector chain cannot be upgraded for LHC deuteron-ion operation without
major hardware upgrades and investments (in equipment and manpower).
Further detailed investigations   are needed to confirm whether
either of the schemes outlined above is feasible.  In any case,
deuterons in the LHC will require several years' lead time.

\subsection{LHC Main Rings }\label{sec:LHCMainRings}

The LHC differs from RHIC in its two-in-one magnet
design, a single magnet ring with two beam apertures, rather than the
two rings of independent magnets of the Brookhaven machine.   With
asymmetric beams in the machine this difference is crucial and
determines many key beam parameters and experimental conditions.

For definiteness, we consider the case of protons colliding with lead
ions; the case with other beams is analogous.   The LHC accelerates
protons through the momentum range
\begin{equation}                  \label{eq:pprange}
\QTY{0.45}{TeV}\ {\mbox{(injection from SPS) }}
\le p_{\rm proton} \le
\QTY{7}{TeV}\ {\mbox{(collision)}}.
\end{equation}
Since the magnetic field is equal in the two apertures,
there is a
relation (equal magnetic rigidity)  between the momenta of proton and
lead ion:
\begin{equation}                         \label{eq:equalrigidity}
    p_{\rm Pb} = Q\, p_{\rm proton},
\end{equation}
where  \(Q = Z = 82\) and \(A =   208\)   for fully stripped Pb ions.

While this places many constraints on
\pPb  operation, it does, on the
other hand, simplify some aspects:  the geometry of the beam orbits does
not change at all so there are no complications with separation magnets
(c.f., the movement of ``DX'' magnets to adjust the collision geometry
in RHIC).

The  centre-of-mass energy and central rapidity shift
for colliding nucleon pairs within ions
\(
\left( {Z_1 ,A_1 } \right)
\),
and
\(
\left( {Z_2 ,A_2 } \right)
\)
\begin{equation}          \label{eq:rootsy}
\sqrt{\sNN}  \approx 2c\,p_{\rm proton} \sqrt {\frac{{Z_1 Z_2 }}{{A_1 A_2 }}}
,\quad \quad
\yNN  \approx  \frac{1}{2}\log \frac{{Z_1 A_2 }}{{A_1 Z_2 }}
\end{equation}
are  direct consequences of the two-in-one magnet design  via
\eqref{eq:equalrigidity}; see also Table~\ref{tab:rootsy}.

\begin{table}
   \begin{center}
   \begin{tabular}{|r|c|c|c|c|}
      \hline
    & \pp & \PbPb & \pPb & \dPb   \\
    \hline
\(\EN/{\rm{TeV}}\)                & 7 & 2.76 & (7,2.76) & (3.5,2.76)\\
\(\sqrt{\sNN}/{\rm{TeV}}\) & 14 & 5.52 & 8.79 & 6.22   \\
\( \yNN \)                    & 0 & 0 & 0.46 & 0.12   \\
      \hline
   \end{tabular}
   \end{center}
   \caption{Beam energy per nucleon,
        \( \EN \approx  (p_{\rm proton},\, p_{\rm Pb}) c /A \),
        center-of-mass collision energy per nucleon, $\sqrt{s_{\rm NN}}$,
        and  central rapidity shift,  \(  \yNN \),  of colliding nucleon pairs for
        maximum rigidity colliding beams in the LHC;  \(  \yNN \)
        is in the direction of the lighter ion.
  }\label{tab:rootsy}
\end{table}

Because of \eqref{eq:equalrigidity}, the two beams have different
speeds and revolution periods on nominal orbits of the same length.  The
RF systems of the two rings of the LHC are perfectly capable of
operating independently
at the different frequencies required during injection and ramping.
However they must be locked together at identical frequencies
in physics conditions to keep the collision points between bunches from
moving.  This forces the beams onto distorted, off-momentum orbits of
different lengths but identical revolution periods.  The amplitude of
the distortion remains within the limits considered acceptable for the
LHC only for \(p_{\rm proton}>\QTY{2.7}{TeV}/c\).   This  imposes a lower bound on
possible collision energies.

At lower energies, therefore, the beams necessarily have different
revolution periods. Each Pb ion bunch encounters up to 5 or 6 proton
bunches as it traverses one of the straight sections around the LHC
experiments where the two beams circulate in a common beam pipe.  At
injection energy, these encounter points move along the straight
section (in the direction of the proton beam) at a rate of 0.15~m per
turn.  They then disappear into the arcs only to re-emerge a few
seconds later in the next experimental straight section.  As the main
bend field is ramped up, this motion slows down, finally freezing when
the energy is high enough that the RF frequencies can be locked
together.  A re-phasing operation (known as ``cogging'') to peg the
collision points in their proper places may still need to be carried
out (although it may be possible to arrange the timing so that this
takes place in the last part of the ramp).

During injection and ramping the bunches are separated in all the
common sections of the LHC so that they never collide
head-on but nevertheless have some long-range beam-beam interaction.  
It was shown~\cite{ref:EPAC2006} that the separation is sufficient
that the magnitudes of these interactions, expressed either as kicks
or parasitic beam-beam tune-shifts, are very small.  The 
strength of the corresponding  ``overlap knock-out'' 
resonances~\cite{ref:MyersOKO} is also relatively small.
For these reasons, it appears unlikely that
the moving beam-beam encounters will have the severe consequences
experienced in analogous conditions at RHIC (see above) and, earlier
still, at the ISR~\cite{ref:MyersOKO}. However this is a tentative
conclusion requiring more detailed justification.

Moreover, the LHC, unlike RHIC, will have the benefit of four
independent transverse feedback systems, one per plane and per
ring,       with bandwidth high enough to act on individual bunches.
This promises to be a powerful tool in damping any coherent motion
induced by the moving encounters.

The moving encounters might also affect the operation of those beam
position monitors that see both beams.  Some modifications of their
electronics may be necessary in order to implement appropriate signal
gating.

With the fairly conservative assumptions of bunches of 
\(
7 \times 10^7
\)
Pb ions (nominal intensity for the \PbPb mode) colliding with bunches of
\(
1.15 \times 10^{10}
\)
protons (10\% of the nominal \pp mode intensity),
with the usual beam emittances, optics and bunch train structure for
Pb beams, a typical initial peak luminosity would be
\begin{equation}                     \label{eq:typicalLuminosity}
L \approx 1.5 \times 10^{29} {\rm{ cm}}^{{\rm{ - 2}}} {\rm{s}}^{{\rm{ - 1}}} .
\end{equation}
Performance beyond this level might be attainable with, most likely,
higher proton bunch intensity.  However it should be remembered that
 \pPb runs
at the LHC are likely to be rather short, with limited time available to
maximise performance.  On the other hand, by the time a \pPb run is
scheduled, operational procedures ought to be well-established and
smooth. More concrete luminosity projections will
come from deeper studies and, most importantly, initial experience of
running the LHC with proton and nuclear beams.

	For a given luminosity, this choice of the maximum possible
	number of bunches is more likely to generate some cancellation
	among multiple, weaker, moving parasitic encounters.  However
	alternative sets of parameters (e.g., half the number of
	nominal intensity Pb bunches against the same number of
\(
2.3 \times 10^{10}
\) 
proton bunches), would lead to similar luminosity.  These could be of
interest if the total Pb intensity is limited because of collimation
inefficiency (e.g., before the proposed dispersion suppressor 
collimators are installed in the collimation insertions).

The question of switching the directions of the p and Pb beams has
been raised.  This is perfectly feasible.
Clearly the experimental advantages would have to be
weighed against the set-up time (presently hard to estimate) during a
short \pA run. If only one direction is possible, it appears that all
experiments would prefer---or accept---protons in Ring 1 and Pb ions
in Ring~2.

In conclusion, preliminary studies 
since have found no major obstacles, in terms of hardware
modifications or beam dynamical effects, to colliding protons and lead
nuclei in the LHC with adequate luminosity.  However further studies
of the beam dynamics are essential to demonstrate the feasibility of
what will probably be the most complicated mode of operation of the
LHC.


\section{p+A AS A BENCHMARK FOR A+A}

Historically, the benchmark role of \pA (or \dAu at RHIC) collisions
has been essential for the interpretation of the heavy-ion results. At
RHIC, two main examples arise: i) the absence of suppression in the
transverse momentum spectrum of the inclusive hadron
production~\cite{Adler:2003ii,Adams:2003im} proved the jet quenching
hypothesis as the genuine final-state effect at work to explain the
observed deficit of high-$p_T$ hadrons in \AuAu
collisions~\cite{Adler:2003qi,Adler:2002xw}; ii) the moderate
 suppression of the $J/\psi$ at central
rapidities~\cite{Adare:2006ns} contrasts with the stronger suppression
predicted by models extrapolating from SPS data, affecting the interpretation
of the corresponding hot nuclear matter effects in \AuAu. At the CERN SPS, the
experimental data on several \pA systems at different energies are
fundamental for the interpretation of the results on $J/\psi$
suppression in \PbPb collisions~\cite{Abreu:2000ni}.

The nuclear modifications of the production cross sections for hard
processes in \pA com\-pared to p+p collisions are studied here with
special emphasis on those involving large virtualities. Predictions
for cross sections with different degrees of nuclear effects are
collected. These processes are expected to provide key measurements of
the validity of QCD factorization in nuclear collisions as well as
constraints on the nuclear parton distribution functions.

The QCD factorization theorem \cite{Collins:gx} provides a
prescription for separating long-distance and short-distance effects in hadronic
cross sections. The leading power contribution to a general hadronic
cross section involves only one hard collision between two partons
from the incoming hadrons with momenta $p_A$ and $p_B$.  The cross
section can be factorized as \cite{Collins:gx}
\begin{equation}
E_h {d{\sigma}_{AB\rightarrow h(p')}\over d^3p'}
=
\sum_{ijk}\int dx' f_{j/B}(x')
          \int dx\, f_{i/A}(x)
          \int dz\, D_{h/k}(z)\,
          E_h {d\hat{\sigma}_{ij\rightarrow k}\over d^3p'} 
          (xp_A,x'p_B,\frac{p'}{z}) ,
\label{twist2conv}
\end{equation}
where $\sum_{ijk}$ runs over all parton species and all scale
dependence is implicit. The $f_{i/A}$ are twist-2 distributions of
parton type $i$ in hadron $A$ (parton distribution functions, PDFs)
 and the $D_{h/k}$ are fragmentation 
functions for a parton of type $k$ to produce a hadron $h$.

In the nuclear case, the incoherence of the hard collisions implies
that the nuclear PDFs (nPDFs) contain a geometric factor, so that the
hard cross sections are proportional to the overlap between the two
nuclei. The degree of overlap can be estimated experimentally in a
probabilistic approach proposed by Glauber~\cite{Glauber:1970jm}. This
fixes the baseline, i.e. the ``equivalent number of \pp collisions'', $N_{\rm coll}$,
 to which the central \AA cross section measurements are compared, 
 to quantify the effects on such observables of the hot and dense matter. 
 The Glauber model is, however, not a
first-principles calculation and experimental checks of this model are of
utmost importance for the interpretation of the main results expected
in the \AA runs.


\subsection{Nuclear parton distribution functions}

\begin{figure}[h]
\begin{minipage}{0.5\textwidth}
\begin{center}
\includegraphics[width=\textwidth]{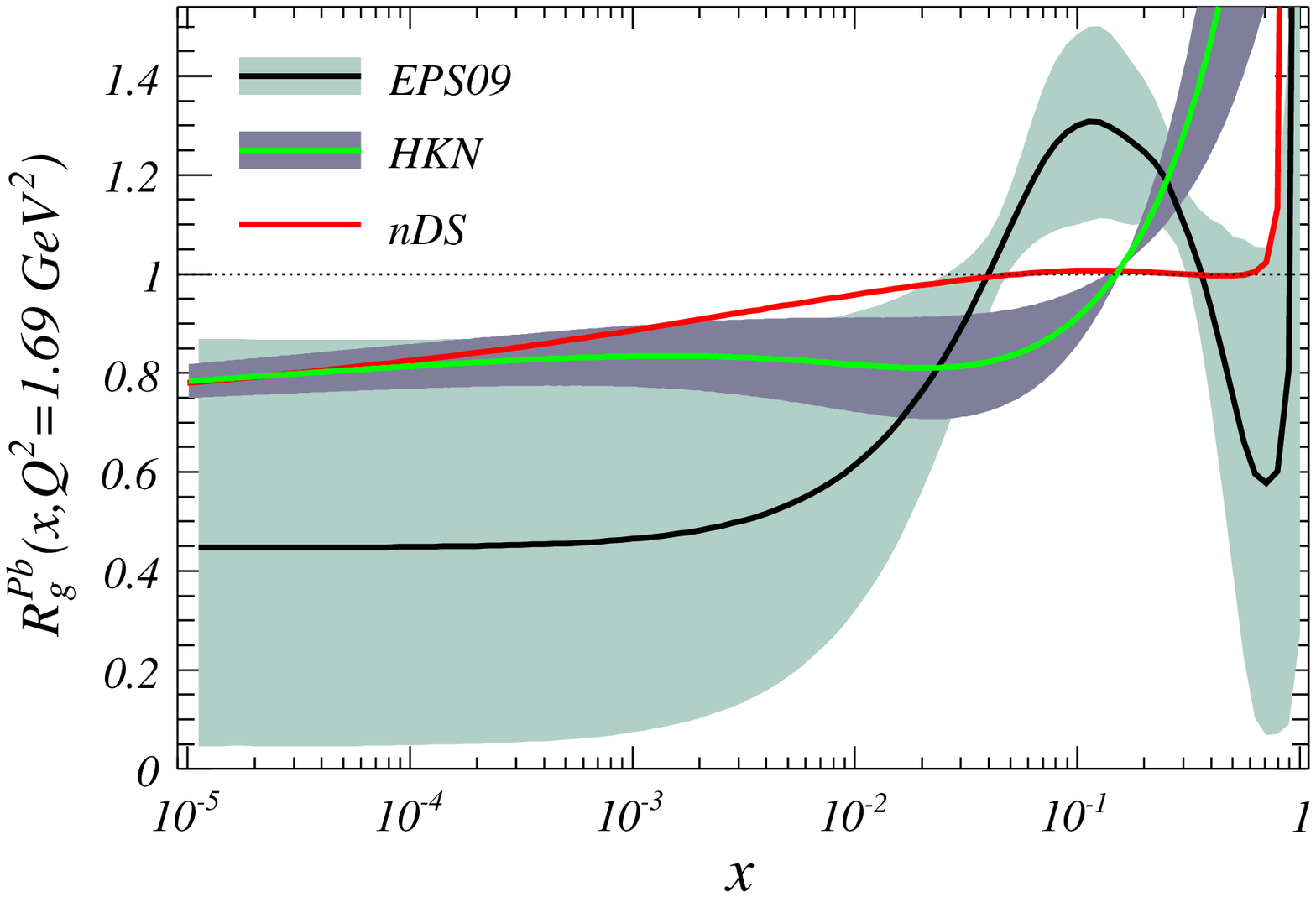}
\end{center}
\end{minipage}
\hfill
\begin{minipage}{0.5\textwidth}
\begin{center}
\includegraphics[width=\textwidth]{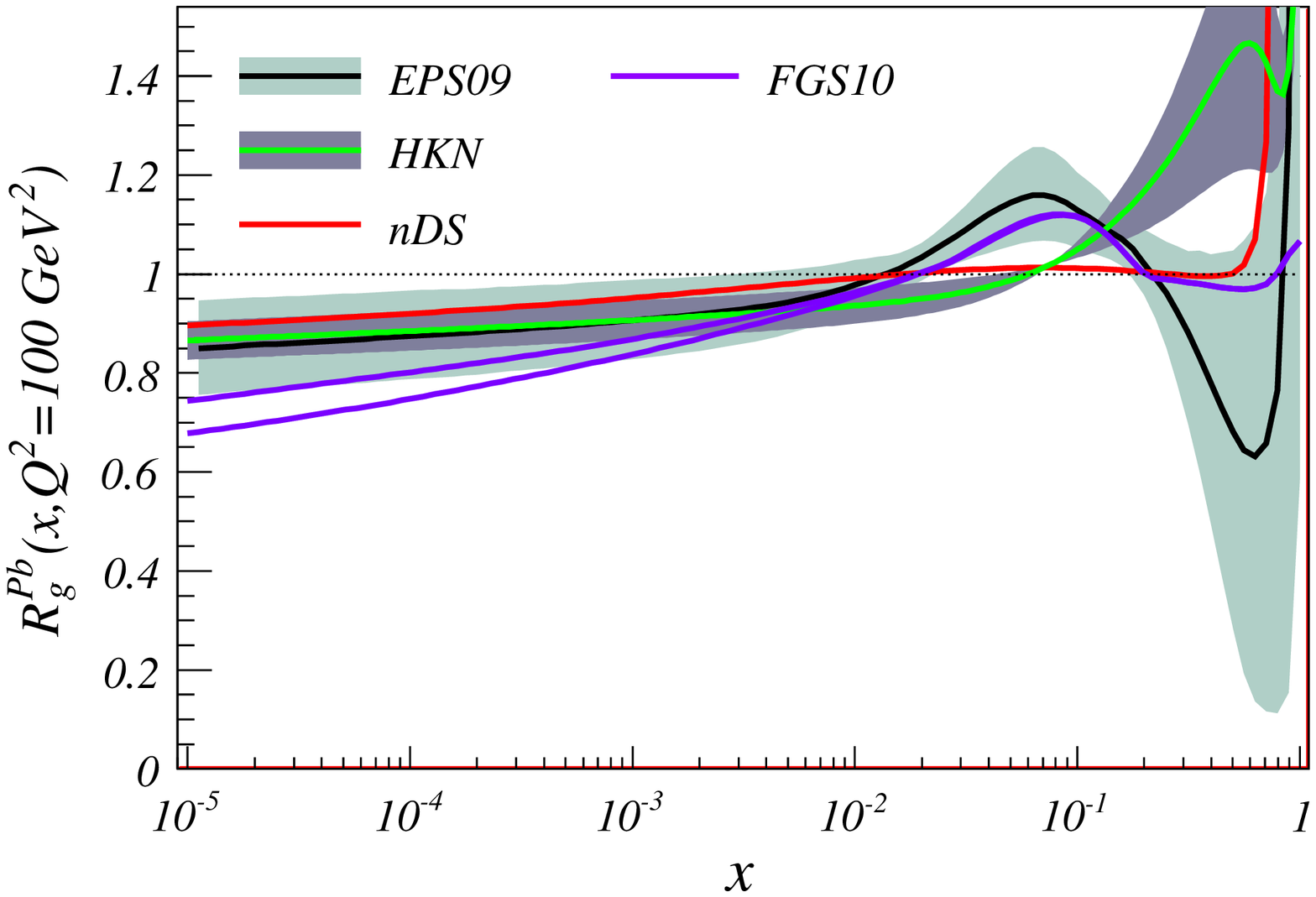}
\end{center}
\end{minipage}
\caption{Current knowledge of nuclear PDFs, shown as the ratio of bound over free proton gluon distributions, $R_g^{Pb}(x,Q^2)$, obtained by the NLO global fits EPS09 \cite{Eskola:2009uj}, HKN07 \cite{Hirai:2007sx} and nDS \cite{deFlorian:2003qf} at two different virtualities, $Q^2=1.69$ GeV$^2$ and $Q^2$=100 GeV$^2$. Also shown for $Q^2=100$ GeV$^2$ are the results from Ref. \cite{Guzey:2009jr} (FGS10) in which gluon shadowing is computed from the DIS diffraction cross section measured at HERA.}
\label{fig:npdf}
\end{figure}

Equation (\ref{twist2conv}) reveals the need for a precise knowledge of the
PDFs for the LHC physics programme. For the proton case, the PDFs are
constrained by a large number of data --- especially from HERA and the
Tevatron --- in global fits performed at LO, NLO or NNLO. In the
nuclear case, much less extensive experimental data on nuclear DIS are available in
the perturbative region ($Q^2\gtrsim$ 1 GeV$^2$), only for $x\gtrsim
0.01$. As a result, there are large uncertainties in the nPDFs relevant for LHC kinematics. The most
recent versions of the nPDFs global fits at NLO are
EPS09 \cite{Eskola:2009uj}, HKN07 \cite{Hirai:2007sx} and
nDS \cite{deFlorian:2003qf} --- also Schienbein {\it et
al.}  \cite{Kovarik:2010uv} performed a similar global fit but did not
release a set for public use yet. Studies of the uncertainties
following the Hessian method are
available \cite{Eskola:2009uj,Hirai:2007sx} and also released for
public use. All sets of nPDFs fit data on charged leptons DIS with
nuclear targets and Drell-Yan in proton-nucleus collisions. Checks of
the compatibility with other hard processes are also available: the
inclusive particle production at high transverse momentum from \dAu
collisions at RHIC has been included in the analysis
 of \cite{Eskola:2009uj} without signs of tension among the different
data sets; the compatibility with neutrino DIS data with nuclear
targets has also been checked in
Ref. \cite{Paukkunen:2010hb}\footnote{See, however,
Ref. \cite{Kovarik:2010uv} for contradicting results.}. Moreover, the
most recent data from $Z$-production at the
LHC \cite{Collaboration:2011ua} also show good agreement with the
factorization assumption although errors are still moderately large. In spite of
these successes, the gluon distribution remains poorly constrained for
the nucleus, as can be seen in Fig. \ref{fig:npdf} where different sets
of nPDFs are shown, together with the corresponding uncertainty bands. 
DGLAP evolution is, however, very efficient
in removing the nuclear effects for gluons at small-$x$, which quickly disappear
for increasing $Q^2$. In this way, these uncertainties become smaller
for the hardest available probes --- see Fig. \ref{fig:npdf} ---
except for the large-$x$ region where substantial effects could
survive for large virtualities. This region is, however, dominated by
valence quarks which in turn are rather well constrained by DIS data
with nuclei.

An alternative approach \cite{Guzey:2009jr} computing the small-$x$ shadowing by its
connection to the hard diffraction in electron-nucleon scattering
has been used to obtain the nuclear PDF at an initial scale
$Q_0$ which are then evolved by NLO DGLAP equations. The inputs in this
calculation are the diffractive PDFs measured in DIS with protons at
HERA. These distributions are dominated by gluons, resulting in a
stronger shadowing for gluons than the corresponding one for
quarks. In Fig.  \ref{fig:npdf} the results from this approach for the
gluon case are also plotted. The differences at small-$x$ become even 
larger at smaller virtualities (not shown)  \cite{Guzey:2009jr}.

It is worth noticing that in contrast to RHIC, where there are constraints at
mid-rapidity ($x\gsim 10^{-2}$) for nuclear distributions from DIS and
DY data, the LHC will probe completely unexplored regions of
phase space. This complicates the interpretation of the \AA data
before a \pA benchmarking programme removes these uncertainties, e.g. for the 
suppression of high transverse momentum particles observed in \cite{Aamodt:2010jd}. 
 The
experimental data from \dAu colli\-sions at RHIC have already proven to
be an appropriate testing ground for nPDFs studies: as mentioned before, data
on inclusive production at high-$p_T$ has been included in global
fits, providing constraints for gluons; nPDFs are also extensively
used in phenomenological studies of hard probes at RHIC. On the other
hand, the strong suppression found in forward rapidity hadron
data \cite{Arsene:2004ux} has challenged the interpretation in terms
of a modification of PDFs alone. Indeed, a global fit including these
data is possible \cite{Eskola:2008ca} but resulting in a sizable tension
with DIS data. The presence of final-state effects and/or the
inadequacy of the collinear factorization formalism --- and the
corresponding onset of saturation of partonic densities --- are two
possible explanations for new mechanisms at work in this rapidity
range.

Reducing the uncertainties on the initial structure of the colliding
nuclei is extremely important also for central conceptual insights
expected from the LHC, such as the evolution of the system in \AA
collisions from cold nuclear matter to hot partonic matter. 
For all dynamical models of this evolution, knowledge of
the initially-produced particle density is crucial.  Ultimately,
however, this density varies with the uncertainty of the nPDFs and
controlling these uncertainties is a decisive step in addressing one
of the central issues in the dynamics of heavy-ion collisions.

In summary, no other experimental conditions, except
\pA collisions at the LHC, exist or will exist in
due time to pin down the
parton structure of the nucleus in the necessary kinematic regime for the \AA studies.


\subsection{Processes of interest for benchmarking}

The characterization of the medium properties in heavy-ion collisions
is performed through processes which couple to the medium in a
theoretically well-controlled manner. Among these processes, the hard
probes --- e.g.  jets, heavy flavor, or quarkonia --- require good knowledge of the
nPDFs and other cold nuclear matter effects. Soft probes such
as collective flow do not require, in principle, any benchmark as the corresponding
signals have so far not been observed in more elementary collisions --- although
some recent results from CMS \pp collisions  \cite{Khachatryan:2010gv}
admit an interpretation in terms
of collective phenomena. Other
hard processes, which do not involve the strong interaction in the
final state, such as direct photon production or $W$/$Z$ production,
may also serve as benchmarks and as checks of the factorization
hypothesis, Eq. (\ref{twist2conv}). In this section we review the
uncertainties associated with hard processes due to nPDFs. We select
processes involving large virtualities, where the nuclear effects in
the parton densities are expected to be small, and processes involving
smaller virtualities where the effects are larger.


\subsubsection{Jets}

The modification of the spectrum of particles produced at large
transverse momentum, {\it jet quenching}, is one of the main probes
for the properties of the hot and dense matter formed in heavy ion collisions
at RHIC. Some of the most interesting results from the first
year LHC run refer to this
observable \cite{Aamodt:2010jd,Aad:2010bu,Chatrchyan:2011sx}. For
this reason, studies of (multi)jet production in \pA collisions are of
great importance as a ``cold QCD matter'' benchmark.  Jet rates in
minimum bias \pPb 
collisions at the LHC (2.75+7 TeV per nucleon) have
been computed at NLO using the Monte Carlo code
in~\cite{Frixione:1995ms,Frixione:1997np,Frixione:1997ks}, with a
renormalization/factorization scale $\mu=E_T/2$ where $E_T$ is the total
transverse energy in the event, and using the CTEQ6.1M \cite{Stump:2003yu} 
nucleon parton densities. Implementing a fixed-order computation,
this code produces at most 3 jets and contains no parton cascade.  The
precision of the computation, limited by CPU time, and the
uncertainties due to the choice of nucleon and nuclear parton
densities, isospin corrections, scale fixing,$\dots$, together with
the influence of the jet finding algorithm and the possibilities to
explore different nuclei and collision energies, have been discussed
elsewhere~\cite{Accardi:2003be,Accardi:2003gp}.

\begin{figure}[htbp]
\begin{center}
\setlength{\epsfxsize=0.65\textwidth}
\centerline{\epsffile{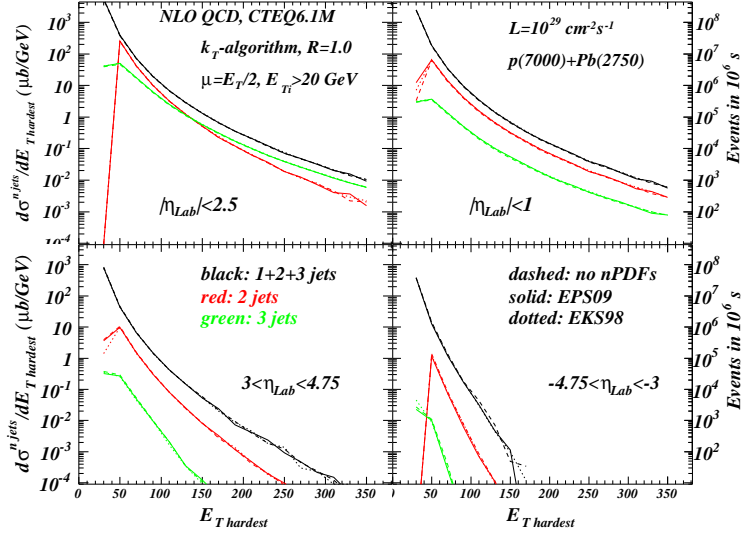}}
\end{center}
\caption[]{1-, 2- and 3-jet cross section as a function of 
the $E_T$ of the hardest jet within the acceptance. Different pseudorapidity windows (in the laboratory frame) computed for minimum bias
\pPb collisions at the LHC (2.75+7 TeV per nucleon) are
considered. Dashed lines are the results without nuclear modification
to the PDFs; solid lines are the results with
EPS09 \protect\cite{Eskola:2009uj}; dotted lines are results with
EKS98 nuclear corrections \cite{Eskola:1998iy,Eskola:1998df} to
nucleon parton densities. Also shown is the scale for corresponding
yields for a luminosity ${\cal L}=10^{29}$ cm$^{-2}$s$^{-1}$ in one
month of running.}
\label{fig:jetppb}
\end{figure}

Figure \ref{fig:jetppb} shows the results for 1-, 2- and 3-jet yields
within two central, one backward and one forward pseudorapidity windows
in the LHC frame (asymmetric for these beam momenta), as a function of the $E_T$
of the hardest jet within the acceptance. The yields, computed here
for a luminosity ${\cal L}=10^{29}$ cm$^{-2}$s$^{-1}$ integrated in
one month ($10^6$ s) run, are quite large --- for simplicity, the
corresponding scale can be read in the two right panels of
Fig.~\ref{fig:jetppb}. For example, in the backward region, $-4.75<\eta
<-3$, yields above $10^5$ 1-jet events per GeV can be achieved for
$E_{Thardest}< 80$ GeV. In the same region, the yields of events with 2
jets within the acceptance are not reduced by than a factor 100. Thus,
studies of cold nuclear matter effects on multi-jet production should be feasible.

The effect of nuclear corrections to PDFs is very small [${\cal O}$(20\%) at most]
and hardly
visible in the yields in Fig.~\ref{fig:jetppb}. The corresponding hot nuclear matter
effects in \PbPb are expected to be much larger. 
 The energy interpolation to make the ratios
with the expectations from \pp without nuclear effects should be safe
enough for the required degree of accuracy.

\subsubsection{Processes involving electroweak bosons}

The production of electroweak bosons has not been studied
in nuclear collisions before the LHC due to the limitations in energy.
However, already
during the first lead-lead run, $Z$ production has been reported by
ATLAS \cite{Collaboration:2011ua} and CMS \cite{:2010px}. At leading
order, the main mechanism of $W/Z$ production is the quark-antiquark
channel and the fact that valence quark distributions are rather well
constrained by nuclear DIS at large-$x$ makes this probe a good one
for constraining the sea quark distributions 
\cite{Vogt:2000hp,Paukkunen:2010qg}. In fact, the
asymmetrical nature of \pA collisions provide an excellent opportunity
for nuclear PDF studies \cite{Paukkunen:2010qg}.

On the other hand, the increasing relevance of jet physics in
heavy-ion collisions render $Z$+jet measurements of great
importance to improve the jet energy calibration.  The inclusive
 $Z$+1jet cross section is known at next-to-leading order in the
strong coupling both for light and heavy-quark
jets~\cite{Arnold:1988dp,Gonsalves:1989ar,Giele:1993dj,Campbell:2003dd}.

\begin{figure}[ht]
  \begin{minipage}{0.5\textwidth}
        \centering
    \includegraphics[width=\textwidth]{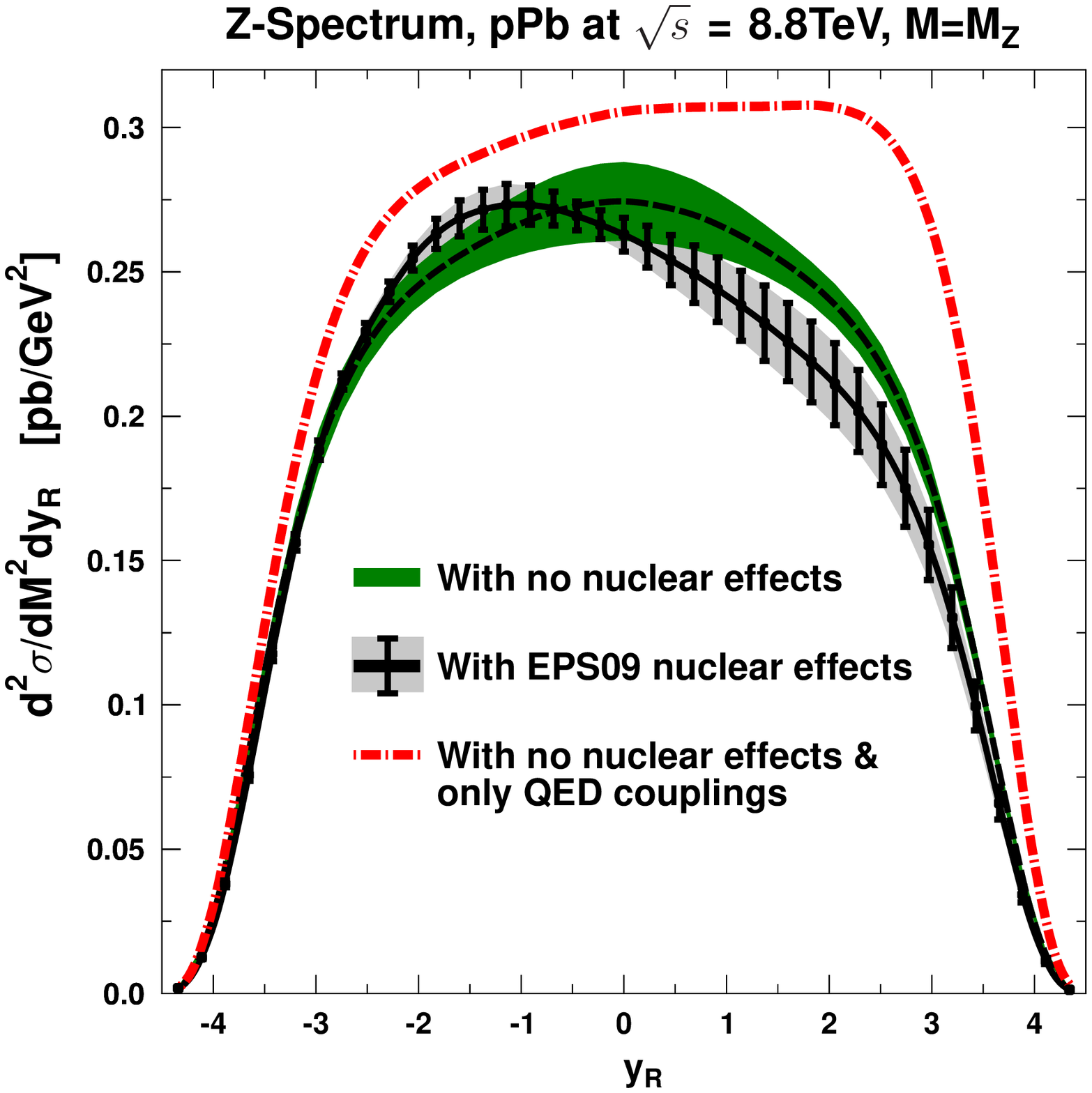}
  \end{minipage}
  ~
    \begin{minipage}{0.5\textwidth}
      \centering
    \includegraphics[width=\textwidth]{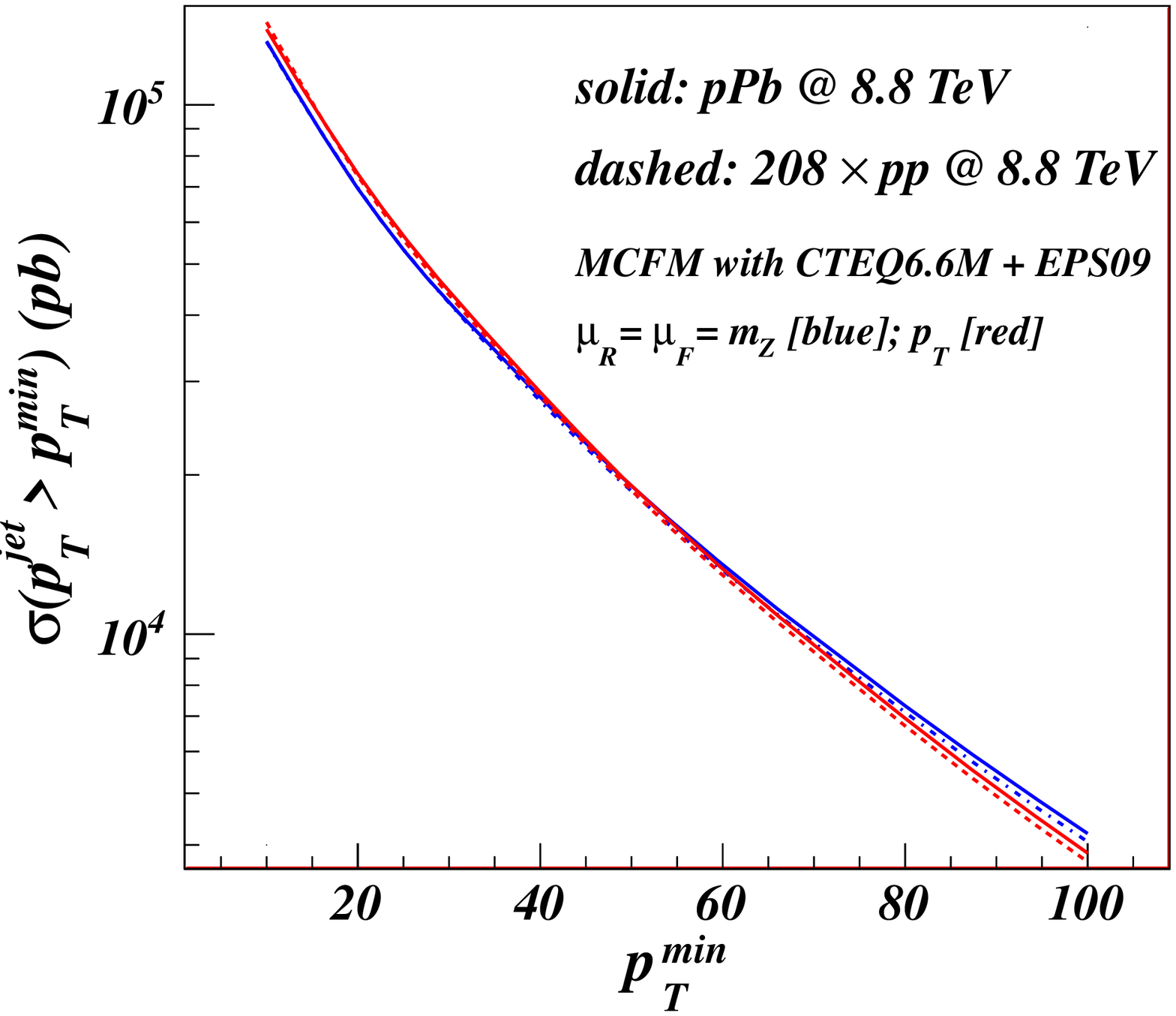}
  \end{minipage} \caption{{\it Left:} Rapidity distributions (in the centre-of-mass
  frame of the \pPb collision) for
  dimuon pairs at the peak of the $Z$ boson in \pPb with and without
  nuclear effects in the PDFs --- the corresponding bands correspond to the 
  uncertainties in the proton PDFs and the nuclear PDFs 
  as given by CTEQ6.6M~\cite{Nadolsky:2008zw} 
  and EPS09~\cite{Eskola:2009uj}
  respectively. For comparison, also the corresponding
  spectrum for only QED couplings (multiplied by a factor 1100) is
  shown by a red line --- Figure
  from \protect\cite{Paukkunen:2010qg}. {\it Right}: Integrated cross
  section $\sigma(pp \to Z [\to \mu^+\mu^-]+$jet) as a function of the
  minimum transverse momentum of the leading jet, $p_T^{\rm min}$,
  with two different scale choices (upper curves at $p_T^{\rm min}=10$
  GeV $\mu_R=\mu_F=p_T$, lower curves
  $\mu_R=\mu_F=m_Z$).}  \label{fig:Zone}
\end{figure}

In Fig.~\ref{fig:Zone} {\it Left} we plot the rapidity distribution
for the NLO production of dimuon pairs at the peak of the mass of the
$Z$-boson in \pPb collisions at LHC energies (notice that the rapidity
refers to the \pPb centre-of-mass frame and cross section is per
nucleon). The fact that the isospin corrections are almost negligible
for $Z$ boson production yields a spectrum which is almost
rapidity-symmetric before nuclear corrections to PDFs are
implemented. This fact provides a clear advantage of the \pPb system
over the \PbPb system as forward-backward asymmetries provide direct
information about the nuclear PDFs without the need for reference \pp
data \cite{Paukkunen:2010qg}.

In Fig.~\ref{fig:Zone} {\it Right} we plot the corresponding NLO
inclusive cross section $\sigma(pp \to Z+$jet), for $Z$ decaying into
leptons, as a function of the minimum transverse momentum
of the leading jet, $p_T^{\rm min}$. The cross section has been
computed with the MCFM package~\cite{Campbell:2000bg} integrating the
dimuon invariant mass region in the range 60 GeV
$<M_{\mu^+\mu^-}<$ 120 GeV. This provides a realistic estimate of the
experimental conditions for measuring the $Z$ mesons at the LHC, as
done, in particular in the corresponding measurement in \PbPb
collisions from Ref. \cite{:2010px}.  Nuclear effects are included
using the parameterization of Ref.~\cite{Eskola:2009uj} and cross
sections are absolute --- for comparison, the \pp cross section at the
same energy and scaled by the corresponding atomic number of the lead
nucleus is also shown.  In this integrated cross sections, the nuclear
effects on the integrated cross sections
were found to be small and are difficult to disentangle from the
typical theoretical uncertainties of the NLO calculation. More
differential distributions are expected to provide further tests of
the nuclear PDFs.

Using the default luminosity quoted in this document, one would expect on the
order of 4000 events with dimuons per unit rapidity in an integrated invariant mass
region around the $Z$ peak at midrapidity --- see
also \protect\cite{Paukkunen:2010qg} --- and a factor of 2 smaller
yields for the case of $Z$+1jet with $p_T^{\rm jet}>10$ GeV. The
corresponding values decrease quickly with increasing $p_T$. For
example, the yield with $p_T^{\rm jet}>60$ GeV is a factor of $\sim10$
smaller. From these results it is clear that a minimum luminosity of $\sim
10^{29}$ cm$^{-2}$s$^{-1}$ is required for these studies
to be feasible. Moreover, in realistic experimental conditions, the
efficiency in the reconstruction of jets would impose a limit on the
minimum $p_T$. A factor of at least 10 more luminosity than the one
reported in the previous section would be a prerequisite for high
enough statistics in $Z$+jet measurements.

It is also worth noting that similar yields are expected in \PbPb
collisions where no extra hot-matter effects are present for the
production of electroweak bosons. The comparison of these two systems
will cross-check the universality of the nPDFs and the
Glauber model as well as precise studies of jet quenching in $Z$+jet
events.


\subsubsection{Photons}

\begin{figure}[ht]
  \begin{minipage}[h]{0.4\textwidth} \centering \includegraphics[width=\textwidth]{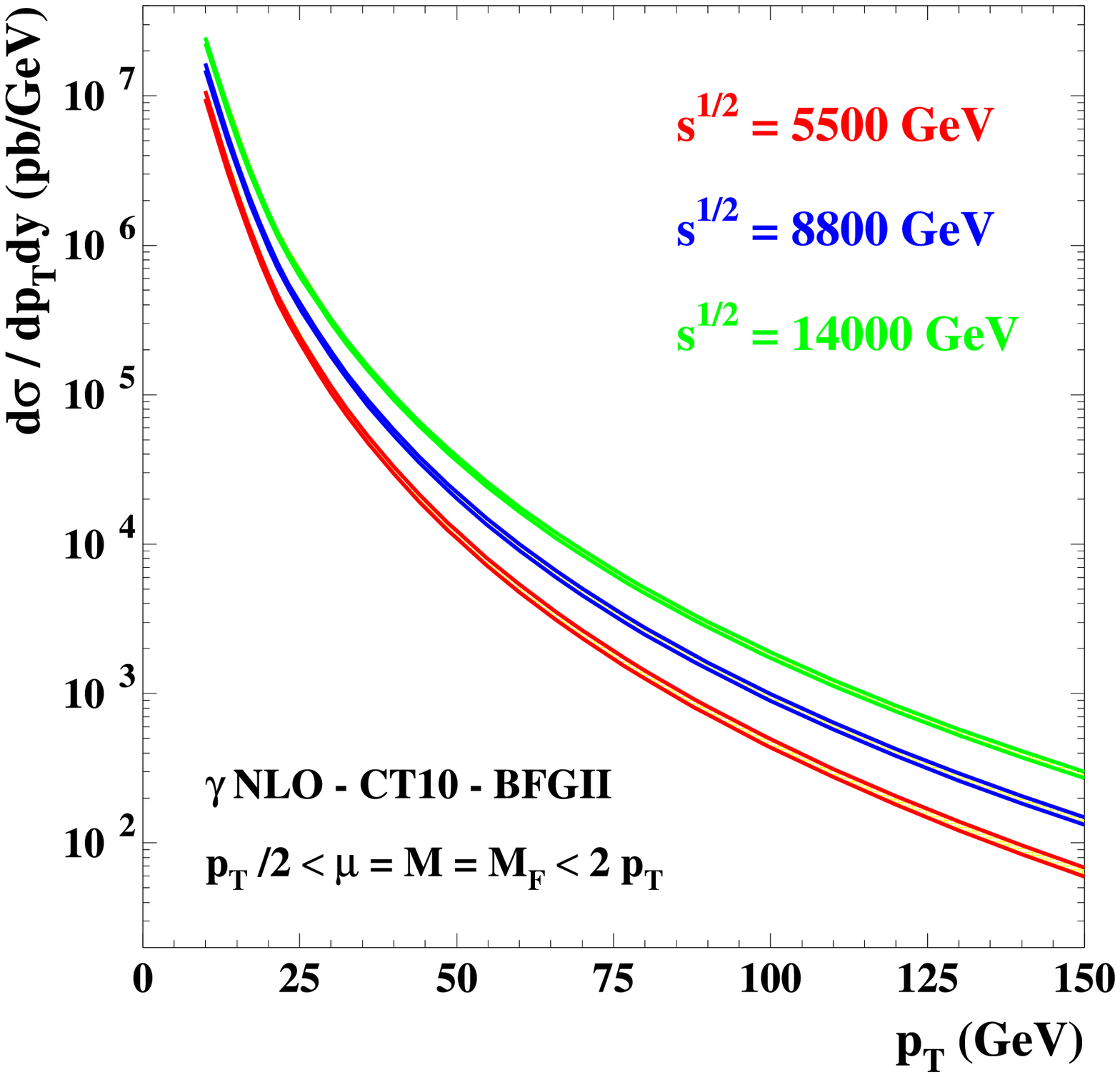} \end{minipage}
    ~ \begin{minipage}[h]{0.6\textwidth} \includegraphics[width=\textwidth]{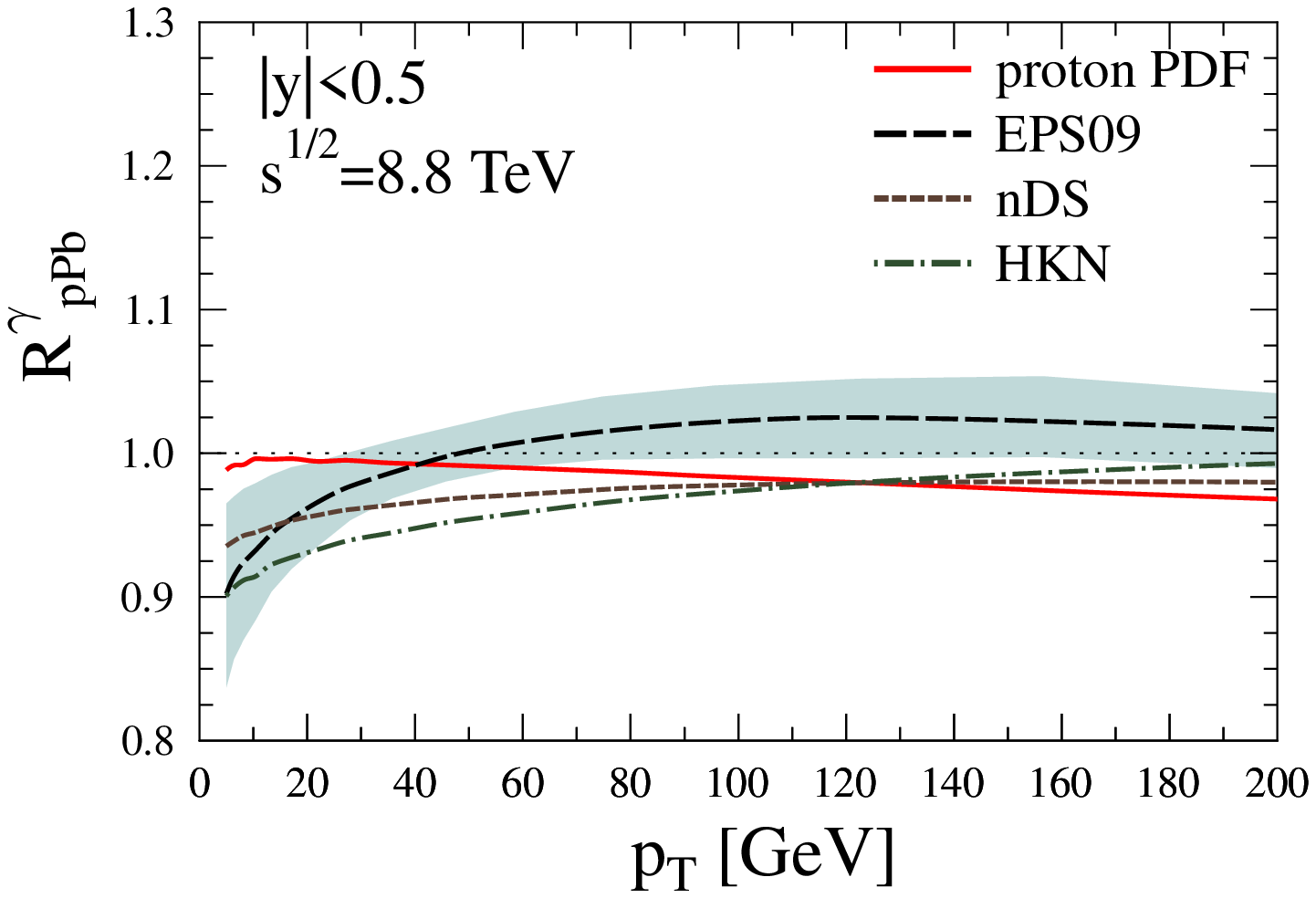} \end{minipage} \caption{
    {\it Left:} Prompt photon $p_{T}$ spectra is computed in
    \pp collisions at $\sqrt{s} = 5.5$, 8.8~and~$14$~TeV. {\it
    Right:} Expected nuclear effects in the scaled ratio of \pPb
    over \pp for different sets of nuclear PDFs,
    EPS09 \cite{Eskola:2009uj}, HKN07 \cite{Hirai:2007sx} and
    nDS \cite{deFlorian:2003qf}. Also shown are the effect of the
    different isospin content of the nuclear and proton
    projectiles. Figure
    from \protect\cite{Arleo:2011gc}.}  \label{fig:spectra}
\end{figure}

Prompt photon production cross sections have been computed in \pp
collisions in QCD at NLO accuracy. We used for the computation the CT10
parton densities~\cite{Lai:2010vv} and the Bourhis, Fontannaz and Guillet
(BFG, set II) photon fragmentation
functions~\cite{Bourhis:1997yu,Bourhis:2000gs}. Fig.~\ref{fig:spectra}
({\it Left}) shows the production cross sections at mid-rapidity 
for \pp collisions at
$\sqrt{s} = 5.5$, 8.8~and~$14$~TeV. The theoretical uncertainties are
estimated by simultaneously varying the renormalization, factorization and
fragmentation scales from $p_{T} / 2$ to $2\, p_{T}$ leading
to a rather stable 20\% systematic error.  

Additional uncertainty should actually come from the rather poorly determined parton-to-photon fragmentation functions \cite{Bourhis:1997yu,Bourhis:2000gs,GehrmannDeRidder:1997gf}.  Although the fragmentation contribution to the photon cross section is about 20\% or less for the fixed target energies, it can easily make up for  about half of the observed photons at collider energies.  An ``isolation'' cut of photon signal can help reduce significantly the less accurate fragmentation contribution \cite{Frixione:1998jh}.  Furthermore, the ``isolation'' cut can help improving the signal-to-background ratio because of the abundance of $\pi^0$'s, which decay into two photons that could be misidentified as one photon at high momentum.

For the production cross sections in \pPb collisions we use the same
setup as in \pp colli\-sions supplemented with different sets of
nuclear PDFs. In Fig. \ref{fig:spectra} ({\it Right}) we present the
nuclear modification ratios for the photon production cross section
in \pPb collisions over that in \pp collisions scaled by the atomic
number of the lead nuclei. Also plotted is the ratio computed with
proton PDFs but including the corrections due to the different quark
content of the neutrons and the protons inside the Pb nuclei (isospin
corrections). The effects are rather small over the entire range of
transverse momentum studied.

As in previous cases, the \pp benchmark for photon production in \pPb
would need an interpolation from the lower energy and the top
energy \pp runs. A potential experimental problem, which could lead to
systematic uncertainties in the comparison, is the 
rapidity shift incurred for asymmetric collision systems.

Inclusive photon production will also be measured in \PbPb collisions
at the LHC. The possible presence of additional hot-matter effects
make the constraints on nuclear PDFs less stringent than in the \pPb
case as the actual size of these effects suffers from large
uncertainties. Turning the argument around, a precise knowledge of the
photon production in \pPb collisions is a necessity to pin down the presence of
additional effects in \PbPb which are not expected to be large.

Finally, we comment on typical counting rates in \pPb
collisions at $\sqrt{s} = 8.8$~TeV assuming a luminosity of ${\cal L}
= 10^{29}$~cm$^{-2}$ s$^{-1}$. While one could expect as many as 1.2
10$^5$ events at $p_{T}$ = 25 GeV in a one month run, this
rate decreases to 5000 events for 50~GeV photon. Still, such a
luminosity would guarantee rather high precision
measurements in the $p_{T}$ range from 25 to 50 GeV.


\subsubsection{Heavy flavor}

The description of heavy quark production in hadronic collisions 
provided by the so-called FONLL (Fixed Order plus Next-to-Leading Log
resummation) approach \cite{Cacciari:1998it} has in recent years been
shown to predict successfully bottom and, to a slightly lesser extent,
charm cross sections in \pp collisions at RHIC and $p+\bar p$ collisions 
at the Fermilab Tevatron. 

For this report we shall restrict ourselves to the small transverse
momentum limit, our main goal being an assessment of nuclear shadowing
effects in \pPb collisions. In this limit FONLL coincides by
construction with the NLO calculation \cite{Nason:1987xz}: it is
therefore the latter which we shall use, complemented with
non-perturbative fragmentation functions identical to those used
in \cite{Cacciari:2005rk}, and the EPS09
parameterization \cite{Eskola:2009uj} of the shadowing effects
implemented in the FONLL package for this purpose. The charm and
bottom mass are set to 1.5 and 4.75 GeV respectively, and the 
CTEQ6.1~\cite{Stump:2003yu}
proton parton distribution functions are employed.

\begin{figure}
\epsfig{file=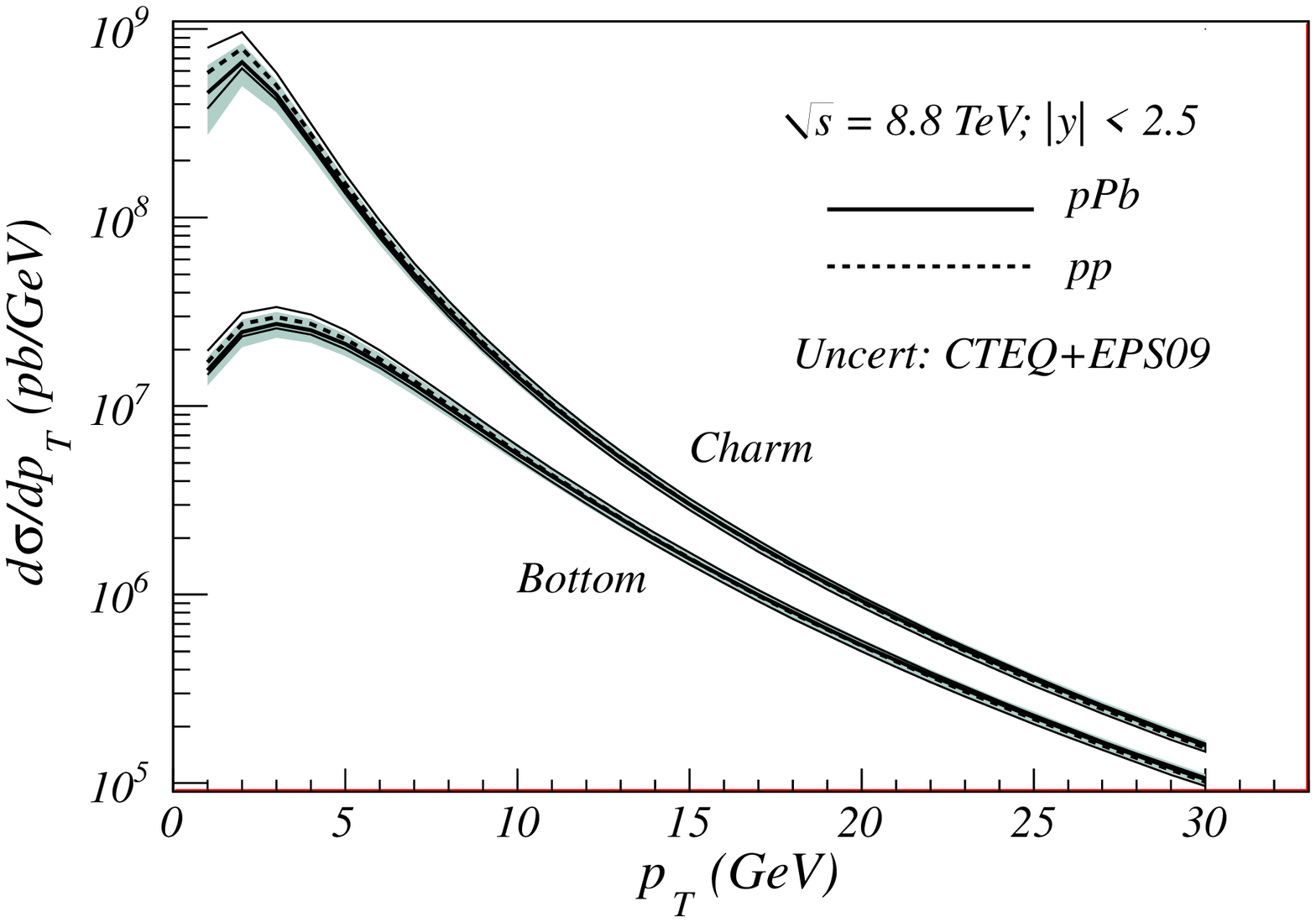,width=0.5\textwidth}
\epsfig{file=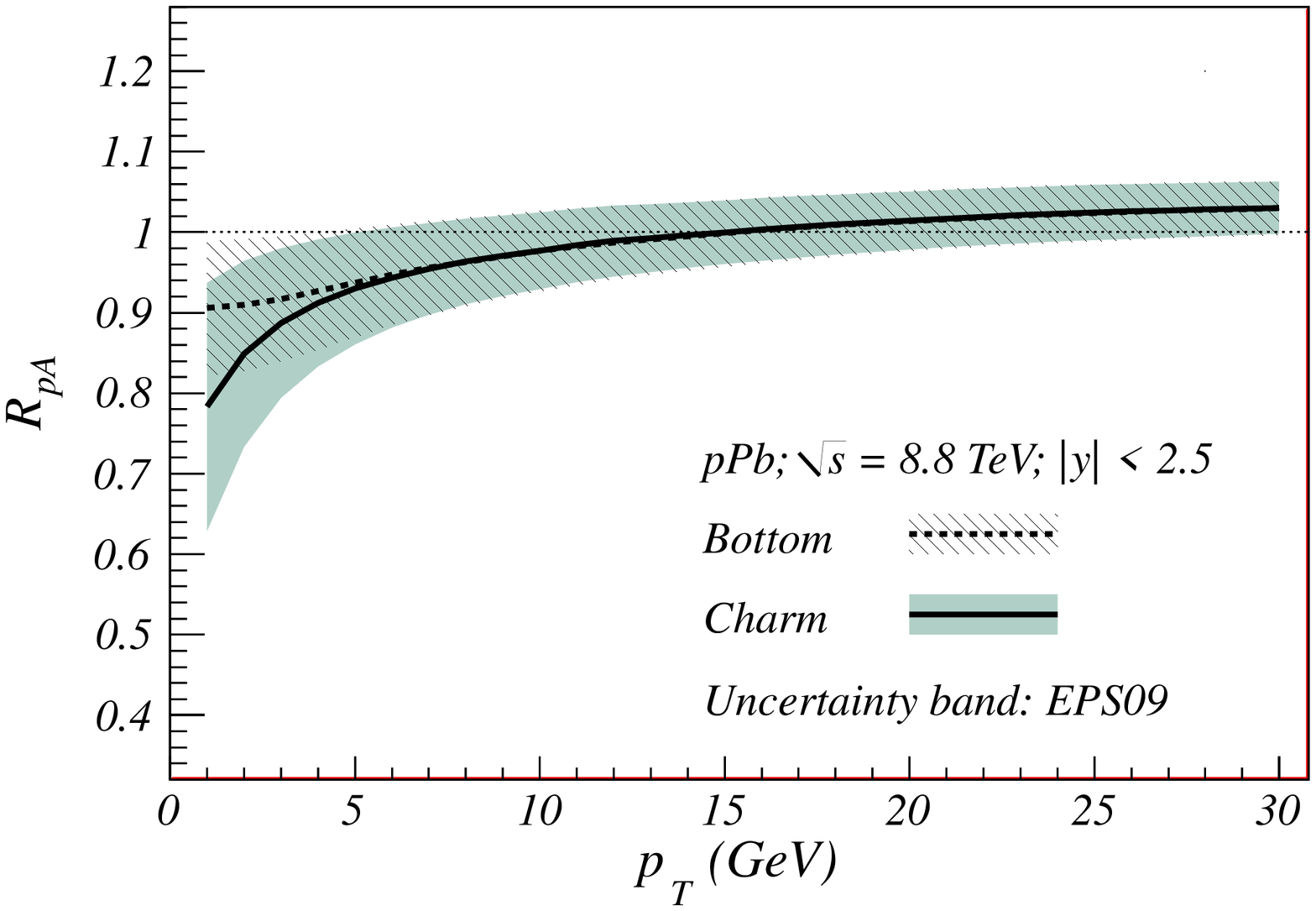,width=0.5\textwidth}
\caption{\label{fig:charm} {\it Left:} Charm and bottom production at the LHC in \pp and \pPb collisions, as predicted by NLO QCD, complemented by EPS09 nuclear corrections where needed. Cross sections are per nucleon and uncertainties refer only to those from the PDFs (proton and nuclear ones added in quadrature). {\it Right:} Nuclear modification factors for both charm and bottom together with the corresponding uncertainties as given by the EPS09 set of nPDFs.}
\end{figure}

The results are shown in Fig. \ref{fig:charm} for both charm and bottom.
 In both cases the transverse
momentum distributions and the nuclear modification ratio $R_{pA}$ are
plotted.  The bands correspond to uncertainties only on the PDFs. In
the case of the total spectra, they include the uncertainties for the
proton and the nuclear PDFs in quadrature, while the ratios include
only those from EPS09 --- this procedure is explained in
Ref.  \cite{Eskola:2009uj}. Although the nuclear effects
are not very large, we note that the associated uncertainties are of the same order
as the effects themselves. Hence, the additional effects expected
in \PbPb collisions would necessitate the \pPb control experiment for a
precise interpretation.

\subsubsection{Quarkonium}

The calculation of quarkonium cross sections in the color evaporation
model is described in Refs.~\cite{Bedjidian:2003gd,Brambilla:2004wf}.
The total yield of lepton pairs from quarkonia decays in \pPb collisions at $\sqrt{s}=8.8$ TeV and
nominal integrated luminosity is $3.9 \times 10^7$ inclusive $J/\psi$
and $2.5 \times 10^5$ inclusive $\Upsilon$ \cite{Bedjidian:2003gd}.
  
We have included intrinsic transverse
momentum, $k_T$, broadening on the quarkonium $p_T$ distributions. 
We found that $\langle k_T^2 \rangle$ value of 2.5
GeV$^2$ is needed for agreement with the Tevatron data.  A simple
logarithmic dependence on the energy,
$\langle k_T^2 \rangle_p = 1 + (1/6) \ln( s/s_0)
\, \, {\rm GeV}^2$
with $\sqrt{s_0} = 20$ GeV, can account for the increase with
increasing $\sqrt{s}$.  Thus for $\sqrt{s}
= 8.8$ TeV, $\langle k_T^2 \rangle_p = 3.03$ GeV$^2$.  
The $k_T$ broadening due to the presence of nuclear matter
is applied as in Ref.~\cite{Vogt:2001nh}. 

\begin{figure}[htbp]
\begin{center}
\setlength{\epsfxsize=0.5\textwidth}
\centerline{\epsffile{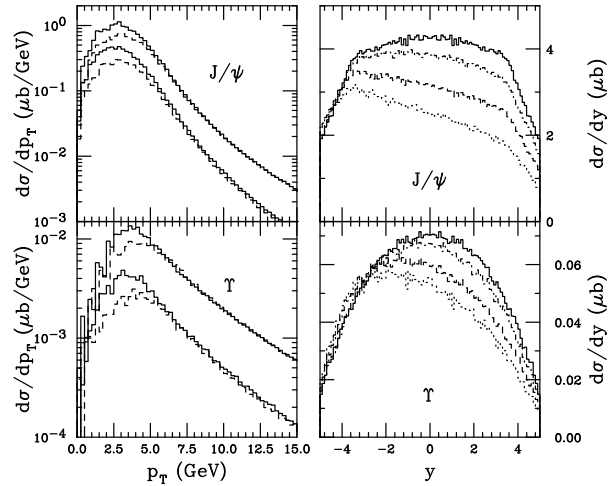}}
\end{center}
\caption[]{The inclusive $J/\psi$ (top) and $\Upsilon$ (bottom) $p_T$ (left)
and $y$ (right) distributions, calculated in the color evaporation model
 for \pp (solid) and
\pPb (dashed) collisions at 8.8 TeV.  The $p_T$ distributions are calculated
using the MRST PDFs and, in \pPb, the EKS98 shadowing parameterization.  
They are shown for the central ($|y|<1$) and forward ($3<y<4$)
rapidity regions.  The forward curves are divided by a factor of 5 for clarity.
The rapidity distributions at 8.8 TeV are shown for the \pp baseline and \pPb
collisions with the nDS (dot-dashed), EKS98 (dashed) and EPS08 (dotted)
shadowing parameterizations. Notice that all cross sections are per nucleon
and rapidity refers to the centre-of-mass frame of the \pPb system.}
\label{fig:totc}
\end{figure}

Sample $J/\psi$ and $\Upsilon$ $p_T$ distributions in 8.8 TeV
\pp and \pPb collisions at the LHC are shown in Fig.~\ref{fig:totc}
in the central region, $|y| \leq 1$ and the forward region $3 < y < 4$.  
The broadening of the $p_T$ distributions in \pPb collisions is rather small.
The effects of initial-state shadowing on the $p_T$ distribution,  
included using the EKS98
parameterization, are likely to be more important.  

The low $p_T$ shadowing
effect on the rapidity distributions can be rather substantial at low $x$.  
High $p_T$ is less affected.  In addition to the EKS98 parameterization in
the dashed histogram, the nDS (dot-dashed) and EPS08 (dotted) shadowing
parameterizations are also shown on the right-hand side of Fig.~\ref{fig:totc}.
Note that at backward rapidity (larger $x$ for the nucleus assuming the
Pb beam moves right to left), the curves tend to coincide with the \pp
curve although the strong anti-shadowing of EPS08 manifests itself for the
$\Upsilon$ at $y < -3$. The deviations from the \pp baseline become stronger
with increasing $y$ (smaller $x$).
The nDS parameterization of the gluon distribution gives the weakest effect
while the EPS08 parameterization is strongest.
(Note that the rapidity shift of the \pPb center of mass
is not shown on the plot.)  For a
complete discussion of the effects of shadowing and nucleon absorption in \dAu
and \pA collisions as a function of rapidity and centrality, 
see Ref.~\cite{Vogt:2004dh}. For recent results on the rapidity dependence of cold matter effects at the
LHC, including calculations with the EPS09 shadowing parameterization, see
Ref.~\cite{Vogt:2010aa}.

From these results it is clear that any conclusion about the effects
observed in \PbPb colli\-sions on quarkonia production need
the \pPb benchmark.

\centerline{   }


In summary, the luminosity quoted in this document is sufficient for
unique studies of perturbative observables. The cases of
electroweak boson or photon production will suffer from a
smaller luminosity. The two main issues to overcome are the
c.m. energy interpolations (between the energies of the A+A and \pA
runs) and the rapidity shifts. For the later, i.e. for those observables
for which acceptance is the limiting factor, more specific studies would
be required and collider operations in the A+p as well as the p+A modes 
are needed.


\section{NEW PHYSICS OPPORTUNITIES: TESTING PERTURBATIVE SATURATION}
\label{subsec:pertsat}

Parton saturation \cite{saturation} is expected to occur at low values
of Bjorken $x$ --- i.e. when the gluon density inside protons and
nuclei becomes large. It can be described by an effective theory
derived from QCD: the Color Glass Condensate (CGC) --- see
e.g. \cite{Gelis:2010nm} for a recent review --- which generalizes the
BFKL evolution equation \cite{Kuraev:1977fs,Balitsky:1978ic} to
situations where the large density of gluons leads to non-linear
effects such as recombination. Searches have been made
for the evidence of parton saturation
effects in small-$x$ data from lepton-proton or
lepton-nucleus collisions as well as in RHIC data from \AuAu and \dAu
systems. Although with the last theoretical developments the
agreement with experimental results is rather good, no firm evidence of the
relevance of saturation physics has been found partly because the usual
DGLAP approaches still provide a very successful description 
of the data also in the small-$x$ region.

Rather general geometrical considerations make saturation effects
larger in nuclei by a factor $\sim A^{1/3}$ as compared to the
proton. As a result, if $x_{\rm sat, p}$ is a typical value at which
non-linear effects appear at a given scale $Q^2_{\rm sat}$ in the
proton, the corresponding value for a nucleus is larger, $x_{\rm sat,
A}\sim A^{\frac{1}{3\lambda}} x_{\rm sat, p}$, for a gluon
distribution behaving as $x^{-\lambda}$. This fact makes nuclear
collisions especially suitable for the study of parton
saturation physics.

The bulk of particle production in nucleus-nucleus collisions has been
computed using methods from the CGC and, in fact, this approach has
been rather successful in predicting the measured \PbPb multiplicities at the
LHC \cite{Aamodt:2010cz}. Heavy-ion collisions, however, are not a
very good testbed if one is interested in the study of saturation
phenomena {\sl per se}. Indeed, in these collisions final state
effects appear which complicate the study of properties
of the wave-function of the incoming projectiles. To
take an extreme view, if the system formed in nucleus-nucleus
collisions reaches a state of local thermal equilibrium, then by
definition it has no memory of its early stages beyond inclusive
properties such as the energy density and perhaps some long-range
correlations in rapidity.

The cleanest experimental situation to look for saturation physics
would be in nuclear DIS experiments at the highest possible energies.
There, one would have direct access to the small-$x$ region of phase space. HERA
experiments so far provide the smallest values of $x$ with protons,
$x\gtrsim 10^{-5}$ for $Q^2\gtrsim$ 1 GeV$^2$, while nuclear data
reaches $x\sim 10^{-2}$ at most. New proposals such as the EIC or the
LHeC~\cite{LHeC,EIC} could extend these ranges significantly. However,
there will be
no overlap in time with the LHC nuclear programme, at least not in the coming
next ten years.  Therefore, proton-nucleus collisions at the LHC offer
a unique opportunity to study the physics of gluon saturation. The
smallest possible values of $x$ in nuclei can be studied at forward
rapidities and with final states that have a moderate transverse mass.

Several different observables have been proposed as good probes of the
saturation of partonic densities. In most of them only one universal
object appears, the so-called ``dipole cross-section''. This
universality can be checked by different measurements at the LHC and
by comparing to smaller energies, in particular with RHIC and with
HERA data.

\begin{figure}[h]
\begin{center}
\includegraphics[width=0.6\textwidth]{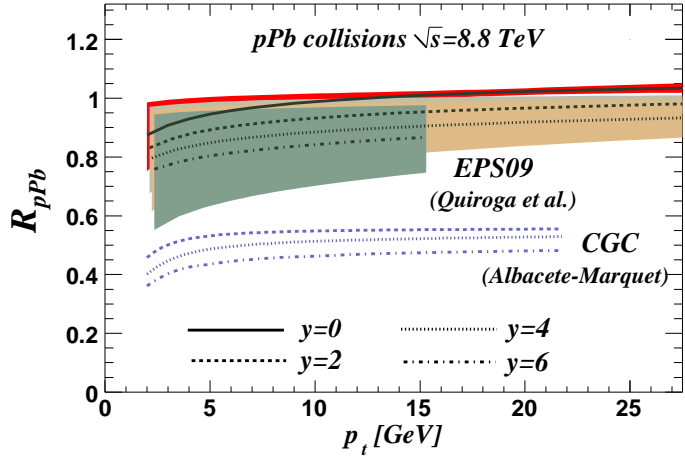}
\end{center}
\caption{Nuclear modification factor for inclusive charged hadrons in p+Pb collisions for different rapidities --- notice, centre-of-mass rapidities here --- computed in the saturation approach of  \cite{Albacete:2010bs} compared with the same quantity computed in collinear factorization \cite{QuirogaArias:2010wh} using EPS09 nuclear PDFs \cite{Eskola:2009uj}, including the corresponding uncertainty bands --- notice that they overlap in most of the $p_T$-region plotted. As explained in the text, the total normalization of the CGC calculation is proportional to the $\langle N_{\rm coll}\rangle$ computed within the Glauber model. For this particular case, this quantity has been fixed to the one computed by the BRAHMS experiment for RHIC conditions, $\langle N_{\rm coll}\rangle = 3.6$.  }
\label{fig:sat1}
\end{figure}

Here, we shall not review the different predictions expected from the
saturation of parton densities. A general effect is that the
presence of non-linear terms in the evolution equations diminish the
growth in the corresponding observables relative to the linear
case. Another generic property of the present implementations is a
correspondence between the rapidity- and $\sqrt{s}$-dependencies
of the non-linear effects which will be testable in \pA collisions at
the LHC. Naively, the effects at $y_{\rm LHC}\sim 0$ are expected to
be similar to those at $y_{\rm RHIC}\sim 3.5$. In order to visualize
this fact, we compare in Fig. \ref{fig:sat1} the nuclear effects in
inclusive hadron production computed in collinear factorization 
\cite{QuirogaArias:2010wh} with a
calculation which uses dipole cross sections evolved with non-linear
Balitsky-Kovchegov (BK) equations including running coupling
effects \cite{Albacete:2010bs}. The last framework is able to reproduce
RHIC data at forward rapidities. Although the BK approach used here is
expected to break at a certain value of the transverse momentum, the
large differences between the two predictions and the large $p_T$-range
available at the LHC will allow to identify a window where the two
scenarios could be cleanly discriminated. It is worth noting here that, at
variance with the collinear factorization approach where the hard
cross section is computed, the CGC approach provides only the spectra
of produced particles. This limitation can be traced back to the fact
that dipole amplitudes are distributed in transverse position and
need to be integrated while this integration is implicit in the PDFs
obtained in the collinear factorization. In this situation, the
computation of the ratios needs information external to the theoretical
framework of the CGC about the
inelastic \pp and \pA cross sections, or, alternatively, about the
average number of nucleon-nucleon collisions $\langle N_{\rm coll}\rangle$, 
usually computed in the Glauber model. This additional ingredient translates
into a normalization factor in the ratio $R_{pA}(p_T,\eta)$ --- 
$\langle N_{\rm coll}\rangle$
 has been fixed to be the same as for RHIC in Fig. \ref{fig:sat1} but
could be larger at the LHC. If the
presence of saturation effects turns out to be a matter of precision,
e.g. compatibility of different data sets within a global fit in
either a DGLAP or a CGC approach, a good control over the
normalization cross sections and/or the validity of the Glauber model
is needed.

%

Another generic feature of the presence of saturation of partonic
densities is the modification of particle correlations due to
collective effects in the initial wave function. In the extreme case,
the momentum imbalance of one given particle can be shared among a
large number of other particles, leading to a loss of the correlation 
signal. Preliminary data from RHIC \cite{Braidot:2010ig} find such a
decorrelation, compatible with a calculation in saturation
physics \cite{Albacete:2010pg}, although alternative explanations have
also been put forward \cite{Strikman:2010bg}.

The present status of the phenomenological calculations within the CGC
framework have strongly benefited from the inclusion of NLO terms in
the corresponding BK evolution equations. These terms are essential to
make the comparison with experimental data meaningful at the
quantitative level and to convert this framework into a predictive
tool for which the \pA programme at the LHC will provide ideal testable
conditions. Improvements in the limitations of the formalism mentioned
above are being worked out and have partly already been used \cite{ALbacete:2010ad}
in the description of the centrality dependence of multiplicities
in \PbPb collisions measured by ALICE \cite{Aamodt:2010cz}.


\section{OTHER OPPORTUNITIES}
\label{sec:opportunities}

\subsection{Ultra-peripheral Collisions}
\label{subsec:upc}

\begin{figure}[tbhp]
  \centering
  \begin{minipage}{0.49\textwidth}
  \includegraphics[width=\textwidth]{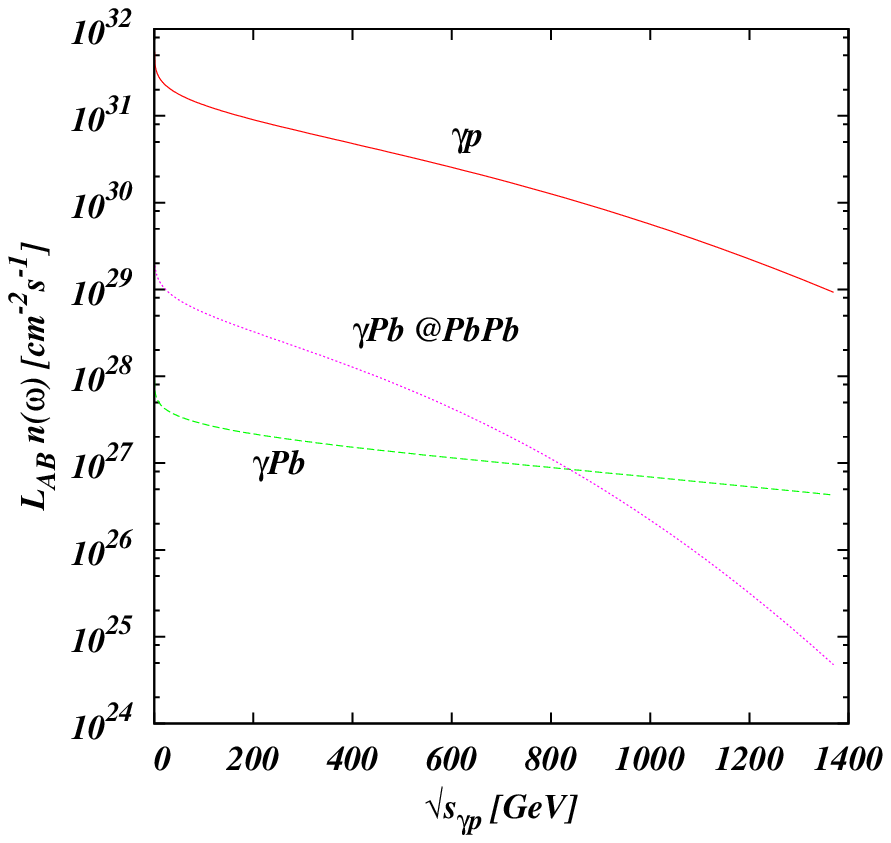}
    \end{minipage}
    \hfill
  \begin{minipage}{0.49\textwidth}
  \includegraphics[width=\textwidth]{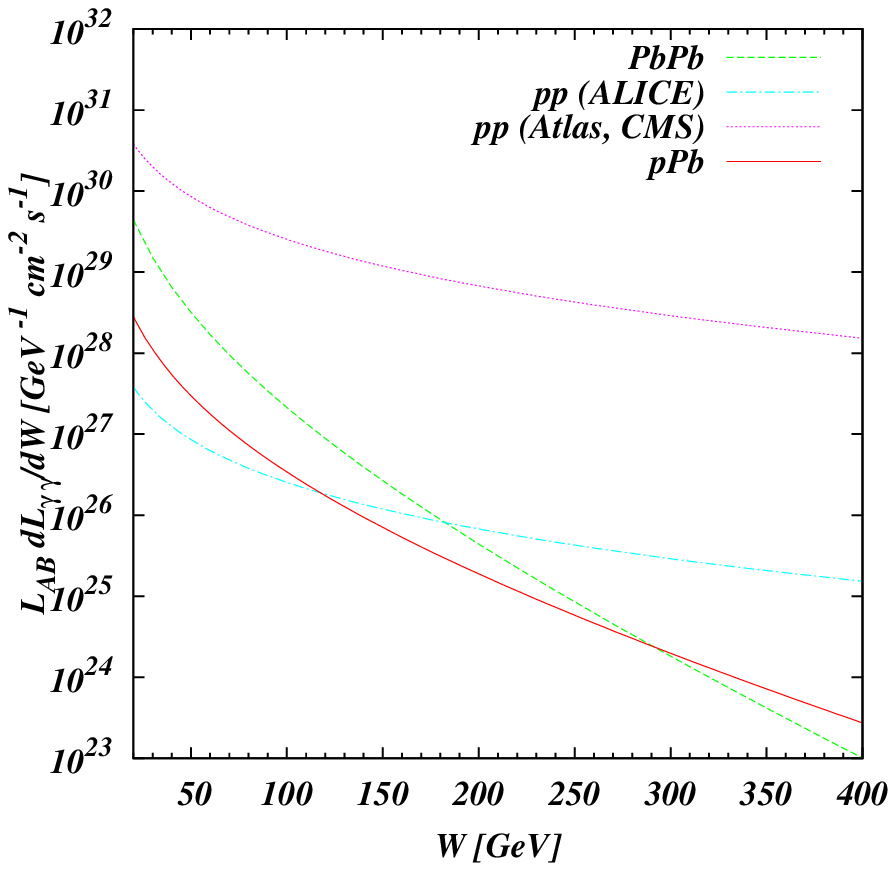}
  \end{minipage}
  \caption{{\it Left:} Effective $\gamma+$A luminosity ${\cal L}_{AB}\,n(\omega)$
    for three cases at the LHC: the photon is emitted from the proton 
(labeled $\gamma$Pb), from the ion ($\gamma$p), and from the ion in a Pb+Pb 
collision ($\gamma$Pb@Pb+Pb).  
    {\it Right:} Effective two-photon luminosities 
  ${\cal L}_{AB}(d{\cal L}_{\gamma\gamma}/dW_{\gamma \gamma})$
    for \pp, p+Pb and Pb+Pb collisions at the LHC. }
  \label{fig:luminosities}
\end{figure}

Charged hadrons accelerated at very high energies generate strong 
electromagnetic fields, equivalent to a flux of quasi-real photons, 
which can be used to study high-energy $\gamma+\gamma$, $\gamma$+p
and $\gamma$+A processes in ultraperipheral collisions (UPCs) where the 
colliding systems pass close to each other without interacting hadronically. 
The effective photon flux, which can be translated into an
effective luminosity, is proportional to the square of the charge, $Z^2$, and
thus significantly enhanced for heavy ions. The figure of merit for
photoproduction is the effective $\gamma$+A luminosity, 
${\cal L}_{AB}\,n(\omega)$, where ${\cal L}_{AB}$ is the accelerator luminosity 
and $n(\omega)$ is the photon flux per nucleus. 
Figure~\ref{fig:luminosities}(a) compares ${\cal L}_{AB}n(\omega)$ for 
$\gamma$+p  and $\gamma$+Pb  collisions in \pPb interactions
to the case where the photon is emitted from an ion in a \PbPb collision.
Figure~\ref{fig:luminosities}(b) compares the same quantity for $\gamma+\gamma$ 
collisions.

UPCs in p+A collisions present advantages with respect to both ultraperipheral 
A+A and p+p collisions. First, relative to A+A collisions, the p+A luminosities 
are three orders of magnitude larger,
and the hadronic center-of-mass energies are also
larger.
Moreover, the
$\gamma+\gamma$ centre-of-mass energies are also higher, resulting from a
harder proton photon spectrum and a smaller distance between the centers 
of the radiating charges.  In addition, it is easier to remove other 
photoproduction backgrounds than in A+A collisions characterized by additional 
photon exchanges which lead to forward neutron emission.  It is also possible
to tag the scattered proton using Roman Pot detectors
(CMS/TOTEM~\cite{CMS-TOTEM-FWD}, ATLAS/ALFA~\cite{Royon:2010sm}, 
FP420~\cite{FP420-LOI}), allowing full kinematic reconstruction by separating 
the momentum transfers from the proton and the ion. 
The advantage of p+A with respect to p+p UPCs is threefold.  First, the photon 
flux of one beam increases by $Z^2$. It is also possibile to trigger on and 
carry out measurements with almost no 
event pileup and also remove most of the exclusive diffractive backgrounds.
Since the nucleus is a fragile object, Pomeron-mediated interactions in p+A 
collisions will, at variance with 
p+p, almost always lead to the emission of a few nucleons detectable 
in the zero degree calorimeters.

The interest in UPCs at the LHC includes QCD studies such as probing the 
low-$x$ gluon distributions in protons and nuclei, beyond the reach of HERA and 
RHIC respectively, 
via inclusive and exclusive dijet, heavy-flavour and vector meson measurements; 
as well as electroweak processes and new physics 
searches~\cite{Baur:2001jj,FELIX,Bertulani:2005ru}. 
An extensive report on the physics of UPCs at the LHC is available 
in Ref.~\cite{Baltz:2007kq}. UPC studies are integral part of the 
ALICE~\cite{Carminati:2004fp,Alessandro:2006yt}, 
ATLAS~\cite{atlas} and CMS~\cite{D'Enterria:2007xr} heavy-ion programmes.  

\subsubsection{Physics potential of photon-proton/nucleus physics}

Ultraperipheral p+A collisions will play a dual role in both extending studies 
of the hadron structure to a new kinematic domain and serving as a reference for similar 
studies in ultraperipheral A+A collisions.  
 
Inclusive photonuclear processes are of particular interest for the study of 
small-$x$ parton densities. Dijet~\cite{Vogt:2004yr}, heavy flavor~\cite{Klein:2002wm}
and quarkonia 
photoproduction can be used to extract small-$x$ gluon densities in protons and 
nuclei. At comparable virtualities, LHC measurements will extend those at HERA 
to an order of magnitude smaller $x$. For example, the $b$ quark rate in p+Pb 
collisions is measurable to $x \sim 10^{-4}$ at $p_T \sim$~5 GeV 
\cite{Strikman:2005yv}. The $c$ quark rate could be measured at even smaller 
$x$ by going to lower $p_T$. 
Comparison of the rapidity dependence of leading charm in $\gamma+$A and 
$\gamma+$p scattering will be a very sensitive test of the onset of the 
nonlinear regime. 

Exclusive photoproduction of heavy quarkonia, $\gamma + A (p) \to V + A (p)$, 
where 
$V = J/\psi, \Upsilon$ and the nucleus A or proton p remains intact) offers a
useful means to constrain the small-x nuclear gluon density down to values 
$x = m_V^2/W_{\gamma +{\rm A,p}}^2$. The mass, $M_V$, and rapidity, $y$, of the 
final-state vector meson can be used to determine the photon energy, $\omega$, in the 
laboratory frame from  $y=\ln(2\omega/M_V)$.  This determination is ambiguous 
in A+A collisions as it is not possible to distinguish which nucleus emitted 
the photon and which was the target, and thus it is only possible to convert 
$d\sigma/dy$, into $\sigma(\gamma + {\rm \, A} \, \rightarrow V + \, {\rm A})$ 
(in the nuclear target frame) for a single photon energy corresponding to $y=0$. 
This ambiguity is avoided in p+A collisions because the photon is most frequently
emitted by the heavier nucleus.  Thus measuring $d\sigma/dy$ as a function of 
$y$ corresponds to measuring the energy dependence of the photoproduction 
cross section.

Exclusive vector meson photoproduction has been studied in UPCs at RHIC by 
STAR~\cite{STARrho,STARda,Abelev:2008ew,Abelev:2009cz}
and PHENIX~\cite{PHENIXupc,Afanasiev:2009hy}.
The LHC p+A measurements would be extremely valuable as a `benchmark' for 
understanding similar UPC A+A data and measuring nuclear shadowing.  
The ratio of
cross sections $\sigma (\gamma + \, {\rm A} \, \rightarrow V + \, {\rm 
A})/[{\rm A}\cdot\sigma(\gamma + \, {\rm p} \, \rightarrow V + \, {\rm A})]$ 
at the same $W_{\gamma +{\rm N}}$
allows for a rather direct measurement of nuclear shadowing.  The A+A data 
provides the numerator while the p+A data are used in the denominator.  Many 
theoretical and experimental systematic uncertainties cancel in the ratio.

Last but not least, the cross section for the electromagnetic proton dissociation reaction
in the field of the nucleus, p+Pb~$\rightarrow$~Pb+X, can be
reliably calculated ($\pm 5\%$ uncertainties) and thus usable as 
``luminometer''. Ultraperipheral d+Au interactions along with electromagnetic
deuteron dissociation have been measured at RHIC, with a cross section of 
$\sigma_{\rm EMD \, dAu}$~=~1.99 b \cite{White:2005kp},
and compared to theoretical predictions~\cite{Klein:2003bz}
to directly determine the luminosity needed to calibrate the 
cross sections of other processes produced in d+Au collisions ~\cite{GSV}.

\subsubsection{Physics potential of two-photon and electroweak processes}

Photon-photon interactions in UPCs at LHC energies can access a rich
QCD, electroweak, and even beyond the Standard Model (BSM) 
programme at the TeV scale. In the QCD sector, 
double vector meson production $\gamma + \gamma \rightarrow V + V$~\cite{FELIX},
will be accessible with similar rates in \pA and \AA collisions.  In addition
heavy flavor meson spectroscopy can distinguish between quark and
gluon-dominated resonances, search for glueballs~\cite{Baur:2001jj,FELIX}, and
study $\eta_c$ spectroscopy through radiative $J/\psi$ decays with larger rates
than in direct $\gamma + \gamma$ production~\cite{Bertulani:2005ru}. 

QED dilepton production is also of interest as a luminosity 
monitor~\cite{Bocian04}.  In CMS~\cite{CMS-TOTEM-FWD}, the CASTOR/TOTEM forward
detectors can measure low-$p_T$ $e^+e^-$ pairs, corresponding to large impact
parameters, where theoretical calculations are most reliable.
Higher-order QED corrections are expected to reduce the huge dielectron
cross sections in Pb+Pb UPCs ($\sigma_{ee}\approx 200$~b). The p+Pb data could 
provide experimental verification of the predicted deviations from the $Z^4$ 
scaling expected for symmetric ion-ion collisions, as yet unobserved at RHIC 
or the SPS.

Tagging forward protons at the LHC with Roman Pots will enhance 
the detection capability of electroweak processes, improving the background 
suppression. 
The addition of far-forward detectors at $\pm 420$~m \cite{FP420-LOI} would improve forward 
light-ion detection, allowing \pA events to be double tagged.
A process well suited to testing the electroweak ($\gamma$WW) gauge boson 
self-interaction is single W photoproduction \cite{Diener} 
from a nucleon in ultraperipheral \pA and \AA~\cite{Baltz:2007hw} collisions. 
Similarly, 10 $\gamma + \gamma \rightarrow {\rm W}^+ + {\rm W}^-$ 
events are expected in a $10^6$ s \pA run. These ${\rm W}^+ {\rm W}^-$ pairs, 
characterised by small pair $p_T$, are sensitive to the quartic gauge boson 
couplings. Lastly, even Higgs boson production would be measurable if the \pA 
luminosity were increased by a factor of 60~\cite{d'Enterria:2009er}.

\subsection{Measurements of Interest to Astroparticle Physics}
\label{subsec:cosmics}

Current cosmic-ray data reveal a rapid increase of the average mass
number $\langle A \rangle$ of the cosmic-ray flux --- i.e. a
transition from lighter (p, He,...)  to heavier composition --- in the
energy range around $\sim 10^{15}$~eV in the laboratory frame,
coinciding with a steepening of the cosmic-ray flux.  In the energy
range around $\sim 10^{18}$ eV there are indications that the
composition becomes lighter again, correlated with a hardening of
the spectral slope of the cosmic-ray flux \cite{watson&nagano}. The
most recent data from the Pierre Auger Observatory points to yet
another change back to a heavier composition in the highest energy
range $\sim 10^{19}$ eV \cite{Augerxmaxdata}. A precise determination
of this quantity would have a profound impact on the knowledge of the
sources of the high-energy cosmic rays.

Direct measurements of the cosmic-ray mass number $A$ are possible
only up to $E < 10^{15}$~eV. Above this energy attempts to infer the
$A$ of the primary particle are based on the measurements of the
extensive air shower (EAS) induced when the cosmic ray interacts upon
entering the atmosphere.  The main source of uncertainty in the
predictions of the EAS observables stems from our limited knowledge of
the features of hadronic interactions in this energy range, in
particular at forward rapidities, where most of the energy of the
shower flows.  In fact, none of the existing hadronic interaction
models currently used for modeling EAS
development \cite{QGSJETII,sibyll,EPOS} is able to provide
a consistent and
satisfactory description of cosmic-ray data due to the unconstrained
extrapolations from energies reached at accelerator based experiments. 
In this
situation, data from the LHC helps to constrain these models and to
improve the interpretation of cosmic ray
measurements~\cite{d'Enterria:2011kw}.  In particular, a
proton-nucleus run would be of utmost importance since the EAS are
predominantly generated in collisions of the cosmic-rays with Nitrogen and Oxygen
nuclei in the upper atmosphere.  The measurements of particle
production at very forward rapidities, accessible to existing LHC
detectors such as the Zero-Degree-Calorimeters~\cite{zdc} and the
LHCf~\cite{lhcf} experiment, are therefore of special importance.


\section{EXPERIMENTAL CONSIDERATIONS}
\label{sec:experiments}

\begin{figure}[h]
\begin{minipage}{0.5\textwidth}
\begin{center}
\includegraphics[width=\textwidth]{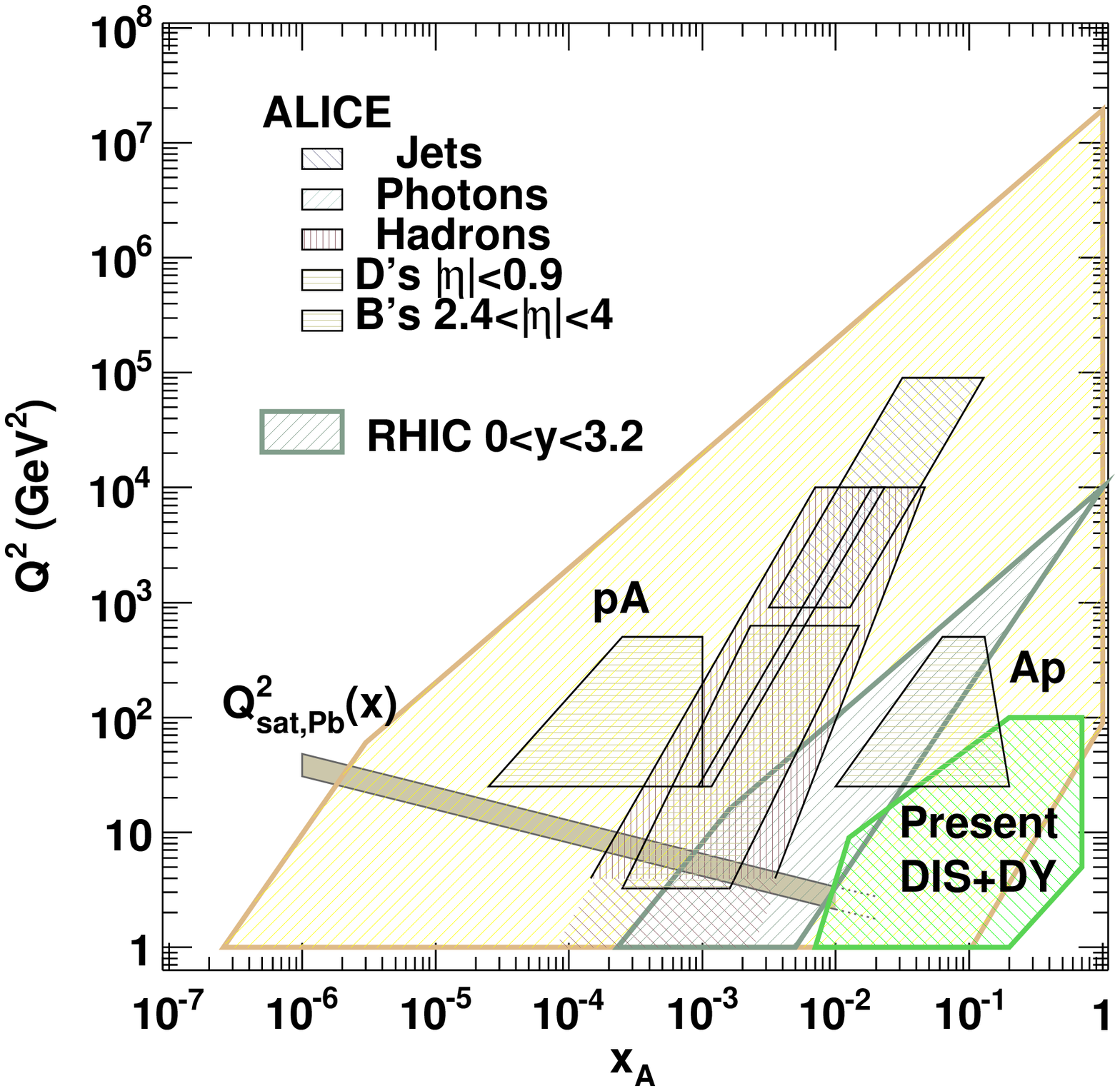}
\end{center}
\end{minipage}
\hfill
\begin{minipage}{0.5\textwidth}
\begin{center}
\includegraphics[width=\textwidth]{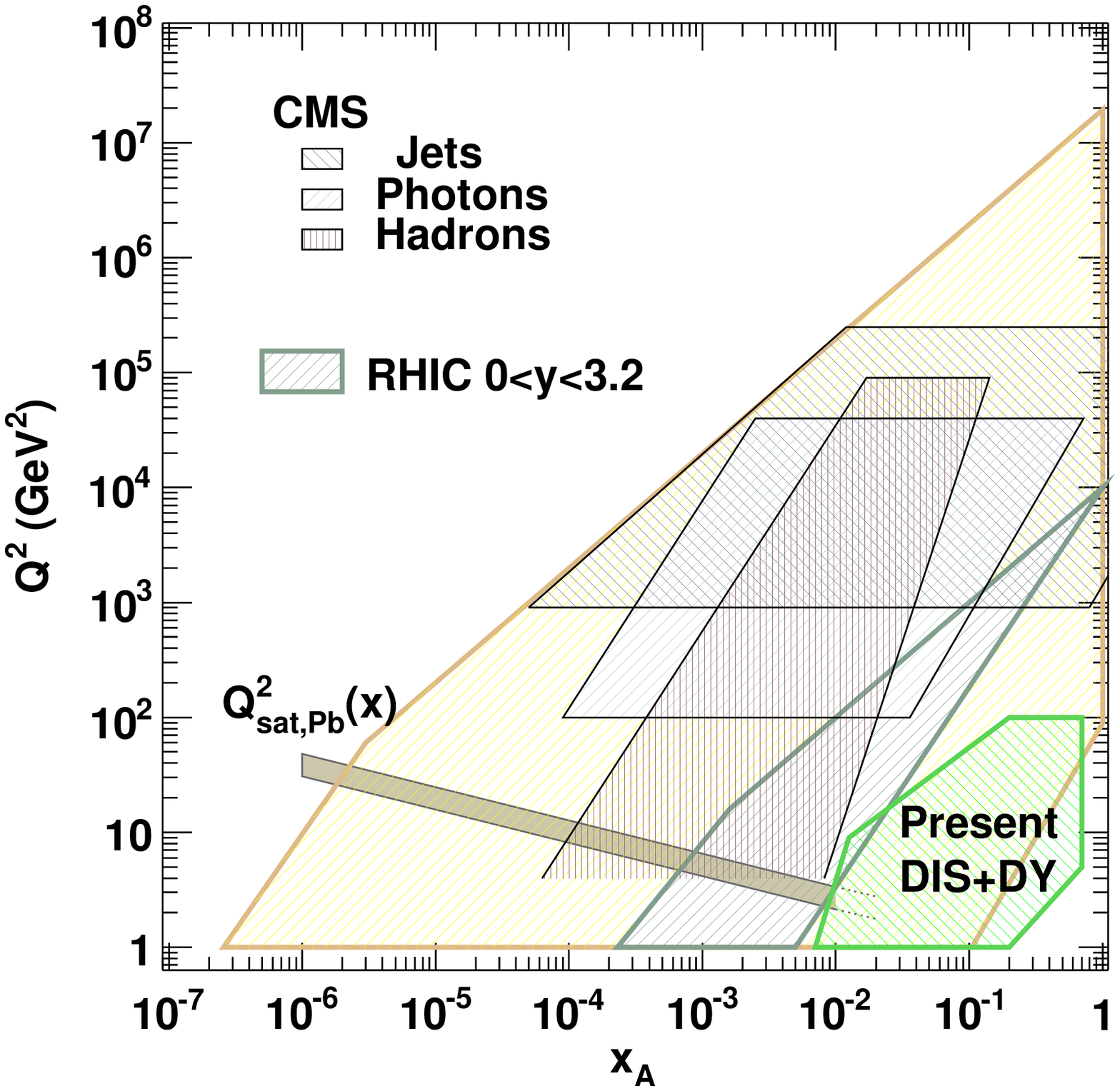}
\end{center}
\end{minipage}
\caption{Total kinematical reach of \pPb collisions at $\sqrt{s}=$8.8 TeV  at the LHC.
Also shown is the reach with an integrated luminosity of
0.1 pb$^{-1}$ for  some of the particular probes studied
in the present 
document for ALICE and CMS, respectively.}
\label{fig:expkin}
\end{figure}

In Fig. \ref{fig:expkin} the expected kinematical regions measured in
the $(x,Q^2)$ plane for different processes\footnote{
These limits correspond to $2\to 2$ or $2\to 3$ processes. If $2\to 1$ kinematics
is relevant, as e.g. in Drell-Yan production or in inclusive hadron production
in the CGC approach, 
the relevant values of $x$ may become significantly smaller.}  accessible with an
integrated luminosity of 0.1 pb$^{-1}$ in a \pPb run are plotted inside
the band indicating the maximum kinematical reach.  Also shown in the same
figure is the reach of the current data used to constrain the present
knowledge of nuclear PDFs. The rest of the phase space to be studied
at the LHC is basically unconstrained with regard to nuclear effects. Notice that the 
band corresponding to RHIC kinematics has to be compared with the
total kinematic reach of the LHC, as the region accessible with actual processes, 
is, in fact, much smaller \cite{Guzey:2004zp}. This
clearly illustrates the wide new region opened at the LHC for both
benchmarking perturbative processes in A+A collisions, and for the new
physics opportunities discussed in this report.

The LHC detectors are designed to run under the much more demanding
conditions of \PbPb and high-luminosity \pp collisions. Extensive
studies of the detector capabilities and performance, including \pA
runs, are
available \cite{Carminati:2004fp,Alessandro:2006yt,D'Enterria:2007xr,atlas}, 
which include discussion of
 the centrality determination in \pA and performance for various
observables. We focus here on topics related to the
rapidity shifts and the energy interpolations for benchmarking as well
as the need of \pA and \Ap collisions during the same running
period. The luminosity quoted in this document, 
${\cal L}=10^{29}$cm$^{-2}$s$^{-1}$, is expected to be sufficient to
carry out the
proposed measurements and matches the detector capabilities. Assuming an 
effective
running time of $10^6$~s, the total integrated luminosity is 0.1
pb$^{-1}$. Any significant reduction of this quantity would significantly
impair the measurement of some of the observables --- e.g. electroweak
boson ($\gamma$, $W$ and $Z$) production --- for benchmarking or other
physics studies as explained previously in the text. On the other
hand, a larger luminosity would allow one to measure with high enough
precision other relevant observables such as $Z$+jet, important for
jet calibration in jet quenching measurements and for nPDF studies.

Benchmarking in \pA is essential for all
perturbative studies relevant for the \PbPb pro\-gramme at the LHC. The
usual procedure is to study ratios between the \AA or \pA cross
sections with the corresponding \pp one scaled by the appropriate
factor. This is, in principle, the same procedure to be used at the
LHC. 
One of the questions is how to produce such ratios, when
the energy of the collision
is different for different colliding systems, as it is the case for the top 
LHC energies in \pp, \pA and \AA. Ideally, 
benchmarking \pp runs at the same energy have to be done. 
This possibility has been explored
 in a short \pp run at $\sqrt{s_{\rm NN}}=$ 2.76 TeV in 2011 with an integrated
 luminosity of around 20 nb$^{-1}$ for ALICE and about a factor of 10 larger for 
CMS and ATLAS. These luminosities are of the same order as the ones considered
here and should provide the needed conditions for benchmarking without 
theoretical input. 
In the absence of \pp data at the same energy as the \PbPb or the \pPb
runs, interpolations among different energies will be needed through a
ratio of the type~\cite{Arleo:2010kw}
\begin{equation}
\sigma(pp \to X; \sqrt{s_{\rm PbPb/pPb}}) = 
\frac{\sigma^{\rm TH}(pp \to X; \sqrt{s_{\rm PbPb/pPb}})}
{\sigma^{\rm TH}(pp \to X \sqrt{s_{pp}})} \sigma^{\rm EXP}(pp \to X; \sqrt{s_{pp}})\,.
\label{eq:interpolationpp}
\end{equation}
where $X$ depends on the process under consideration\footnote{For a first study with
  experimental data on the relevance of these extrapolations see
  e.g. \cite{Aamodt:2010jd}.}.  However, the accuracy of this
  interpolation is process dependent. Uncertainties are quite small
  ${\cal O}(5\%)$ for the hardest processes, e.g. electroweak boson
  production, but are usually larger otherwise. For example they are 12\%
  (8\%) in charm (beauty) production when extrapolating from
  $\sqrt{s}=14$ TeV down to $\sqrt{s}=5.5$
  TeV \cite{Alessandro:2006yt}. 
  At the time of the \pA run some of these uncertainties could be further
  reduced through constraints arising from the \pp program.

Different observables could need slightly different benchmarking
strategies depending on the theoretical or experimental capabilities to
isolate different nuclear effects if present --- for example
modifications in the hadronization. For those effects solely depending
on a modification of the PDFs in the nuclear environment, the use of
ratios with respect to \pp and the corresponding interpolation
(\ref{eq:interpolationpp}) is strictly speaking not essential, especially
if quantities not depending on the knowledge of the proton PDFs
can be built, see e.g. \cite{Paukkunen:2010qg}. In
general the same consideration applies for factorization in the
fragmentation functions when computing e.g. inclusive particle
production at high transverse momentum. In this case, good control
over the normalization of the cross sections would significantly improve the
comparisons as well as allow for precision checks of the Glauber
model.

In this document a canonical energy for a \pPb run of $\sqrt{s}=8.8$
TeV has been considered. This means that a second interpolation to
the \PbPb maximum energy  will be necessary. The
potential problem for this second interpolation lies on the possible
energy dependence of effects which are not factorizable in terms of
nuclear PDFs. With the present knowledge from RHIC and SPS, this will
be especially relevant for quarkonia production --- in this case
a \pPb run at the same energy of the \PbPb one would be preferred. For
other observables such effects are not expected. For nuclear PDF
studies and for the new physics opportunities quoted in this document,
the highest energy run would be more interesting, also for a reduced
systematic uncertainties, if \pp data at the same energy are not
available. Notice that constraints for PDFs in a given range of $x$ are
stronger if the kinematical limit is not reached.

A second independent effect which needs to be taken into account is
the rapidity shift. This rapidity shift, $\Delta y=0.46$, will be
present irrespectively of the energy of the \pPb run. This is
especially relevant for detectors which are not symmetric in
rapidity. In that case, the ideal running conditions for
benchmarking would be to have both \pA and \Ap collisions during the
same running
period. A fast enough switch between the two modes would, in this
case, be a requirement for a successful run.

The integrated luminosity used in this document for benchmarking
purposes is 0.1 pb$^{-1}$. Any substantial reduction of this quantity
would alter the performance of some of the measurements presented
here, in particular those involving electroweak processes 
and large virtualities as mentioned above. A luminosity larger by a
factor of $\sim 10$, on the other hand, 
would give access to new important observables.

\subsection*{Acknowledgements}

The work of  C.~A.~Salgado and N.~Armesto was  supported by Ministerio de
Ciencia e Innovacion of Spain (grants FPA2008-01177 and FPA2009-06867-E), Xunta de
Galicia (Conseller\'\i a de Educaci\'on and grant PGIDIT10PXIB 206017PR), and by project Consolider-Ingenio 2010 CPAN
(CSD2007-00042) and Feder. CAS is a Ram\'on y Cajal researcher.

J.~Alvarez-Mu\~niz thanks Xunta de Galicia (INCITE09 206 336 PR and 
Conseller\'\i a de Educaci\'on); Ministerio de Ciencia e Innovaci\'on 
(FPA 2007-65114 and Consolider CPAN) and Feder Funds.

D.~d'Enterria acknowledges support by the 7th EU Framework Programme (contract FP7-ERG-2008-235071).

Authored by Jefferson Science Associates (V.~Guzey) LLC under U.S. DOE Contract No.
DE-AC05-06OR23177. The U.S. Government retains a non-exclusive, paid-up,
irrevocable, world-wide license to publish or reproduce this manuscript
for U.S. Government purposes.

The work of P.~Jacobs and S.~Klein was funded in part by the U.S. Department of Energy under contract
number DE-AC-76-00098.

The work of J.W. Qiu was funded in part by the U.S. Department of Energy under contract number DE-AC02-98CH10886.

The work  of M.~Strikman  has been supported by the U.S. Department of Energy grant number  DE-FG02-93ER40771.

The work of R.~Vogt was performed under the auspices of the
U.S. Department of Energy by Lawrence Berkeley National Laboratory
under Contract DE-AC02-05CH11231, Lawrence Livermore National Laboratory under
Contract DE-AC52-07NA27344 and was also supported in part by the National
Science Foundation Grant NSF PHY-0555660.

\end{document}